\documentclass[apj]{emulateapj}
\usepackage[colorlinks,linkcolor={blue},citecolor={blue},urlcolor={red}]{hyperref}
 \usepackage{epsfig} \usepackage{graphicx}
\usepackage{natbib} \usepackage{amsmath} \usepackage{amsfonts}
\usepackage{amssymb}

\newcommand{\myemail}{\email{wlixin@shao.ac.cn, leech@shao.ac.cn}}

\newcommand{\mstar}{M$_\ast$}
\newcommand{\lgmstar}{$\lg($\mstar$/$M$_\odot)$}
\newcommand{\mhalo}{M$_{\rm h}$}
\newcommand{\lgmhalo}{$\lg($\mhalo$/$M$_\odot)$}
\newcommand{\mustar}{$\mu_\ast$}
\newcommand{\lgmustar}{$\lg($\mustar$/$M$_\odot {\rm kpc}^{-2})$}
\newcommand{\vdisp}{$\sigma_\ast$} \newcommand{\gr}{$g-r$}

\newcommand{\wrp}{$w_p(r_p)$}

\slugcomment{Accepted for publication in ApJ}  
\shorttitle{Clustering and halo masses of central galaxies}
\shortauthors{Wang, Li \& Jing}

\begin{document}

\title{Clustering properties and halo masses for central galaxies in
  the local Universe}

\author{Lixin Wang\altaffilmark{1}, Cheng  Li\altaffilmark{1,2},  and
  Y. P.  Jing\altaffilmark{3}} \myemail

\altaffiltext{1}{Shanghai Astronomical Observatory,
  80 Nandan Road, Shanghai 200030, China}
\altaffiltext{2}{Physics Department and Tsinghua Center for Astrophysics,
               Tsinghua University, Beijing, 100084, China}
\altaffiltext{3}{Center for Astronomy and Astrophysics, Department of
  Physics, Shanghai Jiaotong University, Shanghai 200240, China}

\begin{abstract}
We investigate the clustering and dark matter halo mass for a sample
of $\sim$16,000 central galaxies selected from the SDSS/DR7 group
catalog.  We select subsamples of central galaxies on three
two-dimensional planes, each formed by stellar mass (\mstar) and one
of the other properties including optical color (\gr), surface stellar
mass density (\mustar) and central stellar velocity dispersion
(\vdisp). For each subsample we measure both the  projected
cross-correlation function (\wrp) relative to a reference galaxy
sample, and an average mass of the host dark matter halos (\mhalo).
For comparison we have also estimated the \wrp\ for the full galaxy
population and the subset of satellite galaxies in the same group
catalog.  We find that, for central galaxies, both \wrp\ and
\mhalo\ show strongest dependence on \mstar, and there is no clear
dependence on other properties when \mstar\ is fixed.  This result
holds at all the masses considered (\mstar$>2\times
10^{10}$M$_\odot$). This  result provides strong support to the
previously-adopted assumption that, for central galaxies, stellar mass
is the galaxy property that is best indicative of the host dark halo
mass.  The full galaxy population and the subset of satellites show
similar clustering properties in all cases. However, they are similar
to the centrals only at high masses (\mstar$\ga 10^{11}$M$_\odot$).
At low-mass (\mstar$\la 10^{11}$M$_\odot$), the clustering properties
of the full population and the satellite galaxies differ from that of
the centrals:  the \wrp\  increases with both \vdisp\ and \gr\ when
\mstar\ is fixed, and depends very weakly on \mstar\ when \vdisp\ or
\gr\ is fixed.  At fixed \mstar, the \mustar\ shows weak correlations
with clustering amplitude (and halo mass in case of central
galaxies). This is true for both centrals  and satellites, and holds
at all masses  and scales except at $r_p$ below a few Mpc and \mstar$\la
10^{11}$M$_\odot$, where the galaxies with higher surface density
are more clustered. Our results suggest that it is necessary to
consider central and satellite galaxies separately when studying the
link between galaxies and dark matter halos.  We discuss the
implications of our results on the relative roles of halo mass and
galaxy structure in quenching the star formation in central galaxies. 
\end{abstract}


\section{Introduction}
\label{sec:introduction}

In current galaxy formation models, galaxies form at the center of
dark matter halos when  gas  is able  to  cool, condense  and form
stars \citep{White-Rees-78}. At this stage the galaxy is called a {\em
  central galaxy}, and its growth is believed to be tightly linked
with the growth of the host dark matter halo. Indeed, a tight
correlation has been well established, both theoretically 
\citep[e.g.][]{Moster-10, Guo-10}
 and observationally \citep[e.g.][]{Mandelbaum-06, Yang-07, 
Li-12c, Li-Wang-Jing-13}, 
between the total mass of stars in a central galaxy
(stellar mass, \mstar) and the dark matter mass of its host halo
(\mhalo). At  a later stage the central  galaxy, together with its
host  dark halo, may be accreted into a  larger halo, and consequently
the  galaxy becomes a {\em satellite galaxy} and its halo  becomes a
{\em subhalo} of the new system. As a result the galaxy will deviate
to varying degrees from the stellar mass--halo mass relation of the
central galaxies.  In addition, environmental effects such as gas
stripping by ram-pressure \citep{Gunn-Gott-72, Abadi-Moore-Bower-99}
and tidal interactions \citep{Toomre-Toomre-72, Moore-96} occurring
within the new system may effectively reduce the hot gas halo, and in
some cases even the cold gas and stellar components of the satellite
galaxy, thus making the correlation with halo mass even more loose. It
is thus important to consider central and satellite galaxies
separately when studying the physical link between galaxies and dark
halos.

Over the past decade, extensive studies have attempted to determine
the halo mass as a function of stellar mass (or stellar luminosity) for
central galaxies using a variety of observational probes such as
galaxy clustering, satellite kinematics, gravitational lensing and
group/cluster catalogs. These studies have obtained a
\mhalo--\mstar\ relation that follows  a double power-law form with a
small scatter of $\sim0.16$ dex in \lgmstar\ at fixed
\mhalo\ \citep[e.g.][]{Yang-03, Vale-Ostriker-04, vandenBosch-04,
  Conroy-Wechsler-Kravtsov-06,    Mandelbaum-06, Wang-06,   Wang-07,
  Yang-07, Baldry-Glazebrook-Driver-08, More-09,
  Behroozi-Conroy-Wechsler-10,  Guo-10,  Moster-10, Li-12c, More-11,
  Li-Wang-Jing-13,Kravtsov-Vikhlinin-Meshscheryakov-14}.  This
relation has been widely-adopted to populate dark matter halos  with
galaxies in current physical/statistical models such as semi-analytic
models \citep{White-Frenk-91, Kauffmann-Nusser-Steinmetz-97},  halo
occupation distribution models \citep{Jing-Mo-Borner-98, Benson-00},
and subhalo abundance matching models \citep{Vale-Ostriker-04}. 

Besides stellar mass, other galaxy properties such as color, star
formation rate, morphology and internal structure are also found to be
correlated with halo mass, as usually probed by galaxy clustering
\citep[e.g.][]{Li-06b, Zehavi-11}, halo occupation statistics
\citep[e.g.][]{Yang-Mo-vandenBosch-08}, abundance matching
\citep[e.g.][]{More-11}, and weak lensing
\citep[e.g.][]{Mandelbaum-06}. In particular, the SDSS data have
recently revealed that, when both central and satellite galaxies are
considered, the galaxy-galaxy correlation function appears to be less
dependent on galaxy stellar mass, when compared to the parameters
quantifying the prominence of a central bulge in the galaxy, such as
central stellar velocity dispersion (\vdisp;
\citealt{Wake-Franx-vanDokkum-12}).  This phenomenon may be attributed
to the population of satellite galaxies  for which the environmental
processes such as tidal stripping occurring within their halos have
stronger effect on stellar mass  than on central stellar velocity
dispersion. When only considering central galaxies and when stellar
mass is fixed, halo mass shows little dependence on both the
properties indicative of star formation history
\citep[e.g.][]{More-11} and the \vdisp, as recently demonstrated in
\citet[][hereafter L13]{Li-Wang-Jing-13}. In L13, the dark matter halo
mass  was estimated for samples selected by \mstar\ and \vdisp, by
modelling the redshift-space distortion in the group-galaxy cross
correlation function. It was clearly shown that, for central galaxies
in the local Universe, stellar mass is better than stellar velocity
dispersion as an indicator of dark matter halo mass.

In the current paper we extend the work of L13 by considering two more
galaxy properties: optical color index (\gr) and surface stellar mass
density (\mustar).  The \gr\  is a parameter associated with the
recent star formation history of the galaxy, whereas \mustar\ is
related to galaxy structure. We will estimate both clustering and halo
mass for subsamples selected on two-dimensional planes each formed by
\mstar\ and one of the three properties: \gr, \mustar\ and \vdisp.  In
addition to analyzing the central galaxies, we will study the
clustering properties of both the full galaxy population and the
subset of satellite galaxies, and compare  the results for the
different populations.  Through these analyses, we aim to answer the
following questions:
\begin{itemize}
\item Is stellar mass the galaxy property that is most related
  to halo mass? How is stellar mass compared to the recent star
  formation history (as indicated by \gr), the structural properties
  (as quantified by \mustar) and the central stellar velocity
  dispersion (\vdisp) in terms of linking galaxies with dark halos? 
\item Is the correlation of halo mass and clustering with galaxy
  properties always stronger when we consider central galaxies only?
  What are the relative contributions of centrals and satellites to
  the overall clustering of galaxies?
\end{itemize}

The paper is structured as follows. In \S\ref{sec:data} we describe
the galaxy sample and the physical properties used in this work. In
\S\ref{sec:centrals} we focus on central galaxies, presenting the
co-dependence of clustering and halo mass on galaxy properties. In
\S\ref{sec:all} we examine the clustering properties for the full
galaxy population and the satellite galaxies. Finally, we summarize
our results and discuss on the implications on the link between
galaxies and dark matter halos, as well as the relative roles of halo
mass and galaxy structure in quenching the  star formation in central
galaxies. 

Throughout this paper we assume a cosmology model with the density
parameter $\Omega_m = 0.27$ and the cosmological constant
$\Omega_\Lambda = 0.73$, and a Hubble constant $H_0 = 100h$ km
s$^{-1}$Mpc$^{-1}$ with $h = 0.7$.

\section{Data }
\label{sec:data}

\subsection{Samples}

For this work we have constructed three samples: i) a sample of central
galaxies based on the SDSS group catalog of \citet{Yang-07}, ii) a
reference sample of galaxies selected from the SDSS data release 7
\citep[DR7][]{Abazajian-09}, and iii) a random sample that  has the
same selection effects as the reference sample. Here we briefly
describe these samples and refer the reader to our previous studies
for  more details: \citet[][hereafter L12]{Li-12c} and L13.

\begin{description}
  \item[Central galaxy sample:] \citet{Yang-07} constructed a group
    catalog by applying a modified version of the halo-based
    group-finding algorithm of \citet{Yang-05c} to  a sample of
    $\sim6.4\times10^5$ galaxies that are in the redshift range
    $0.01\leq z\leq 0.2$ and selected from {\tt sample dr72} of the
    New York University Value-Added  Galaxy Catalog (NYU-VAGC). The
    NYU-VAGC is a  catalog of local galaxies from the SDSS/DR7,
    publicly available at http://sdss.physics.nyu.edu/vagc/, and is
    described in detail in \citet{Blanton-05b}. Following L12 and L13
    we restrict ourselves to a subset of the group catalog, which
    consists of $\sim16,000$ group systems with the number of member
    galaxies $N_{\rm mem}\ge3$ and  the stellar mass of the central
    galaxy \mstar$>2\times10^{10}$M$_\odot$.  We use the most massive
    galaxy member as the central galaxy of each group.

  \item[Reference galaxy sample:] The reference sample is a
    magnitude-limited galaxy catalog selected from the NYU-VAGC {\tt
      sample dr72}, consisting of about half a million galaxies with
    $r<17.6$, $-24<M_{^{0.1}r}<-16$ and $0.01<z<0.2$. Here, $r$ is the
    $r$-band Petrosian apparent magnitude, corrected for Galactic
    extinction; $M_{^{0.1}r}$ is the $r$-band Petrosian absolute
    magnitude, corrected for evolution and $K$-corrected to its value
    at $z=0.1$. 

    \item[Random sample:] The random sample is constructed from the
      reference sample itself following the method described in
      \citet{Li-06b}. For each real galaxy in the reference sample, we
      generated 10 sky positions at random within the mask of the real
      sample, assigning to each of them the redshift of the real
      galaxy. The resulting random sample is expected to be
      unclustered but fills the same sky region and has the same
      position- and redshift-dependent effects as the real
      sample. Extensive tests in \citet{Li-06b} show that random
      samples constructed in this way are valid for  clustering
      analyses, provided that the survey area is large enough and that
      the effective depth of the survey does not vary.  Both
      requirements are met to good accuracy by our reference sample,
      which covers $\ga 6000$ deg$^2$, complete down to $r=17.6$ and
      little affected by foreground dust over the entire survey
      footprint. 

\end{description}

\begin{figure*}
  \begin{center}
    \epsfig{figure=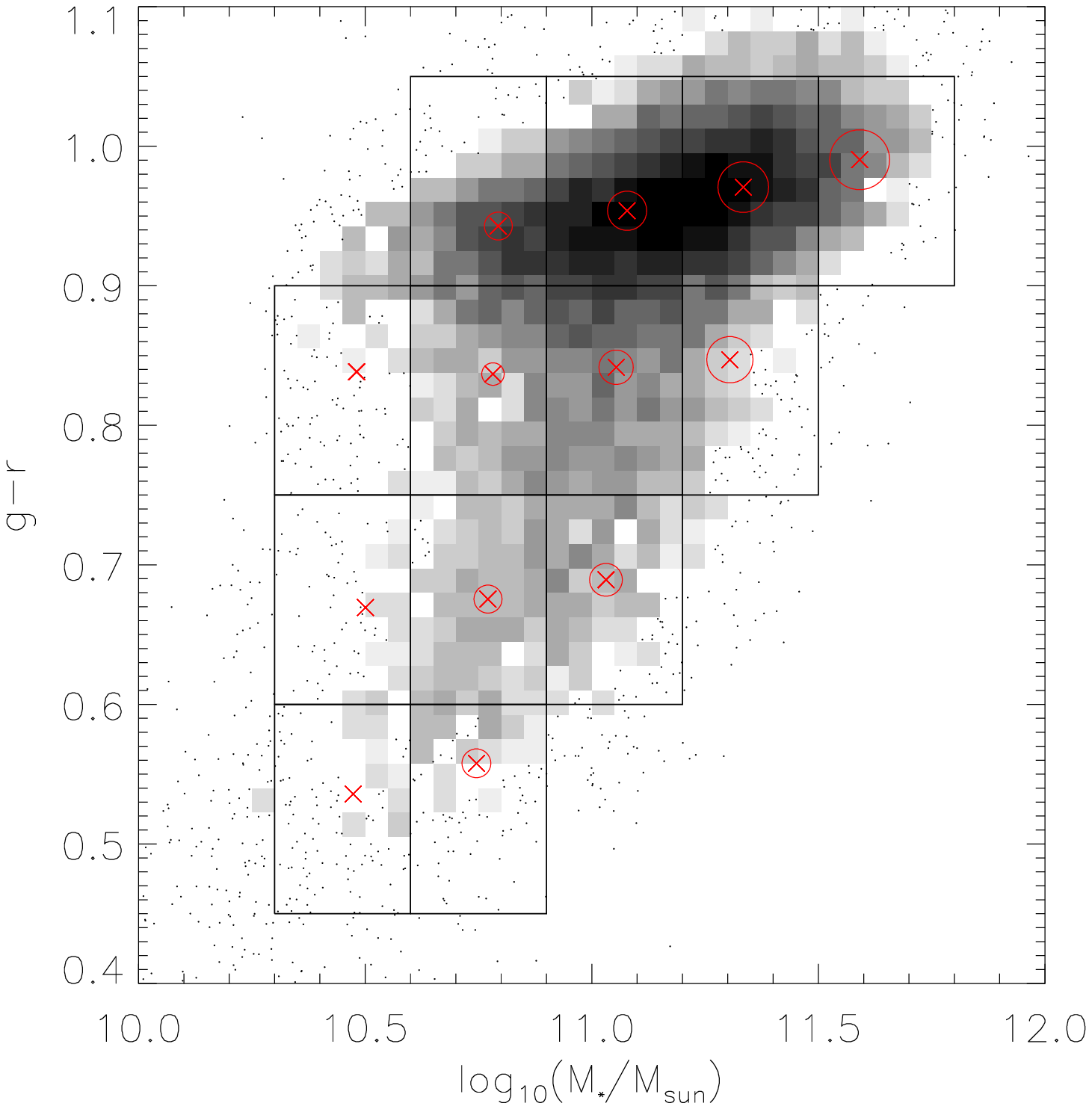,width=0.32\textwidth}
    \epsfig{figure=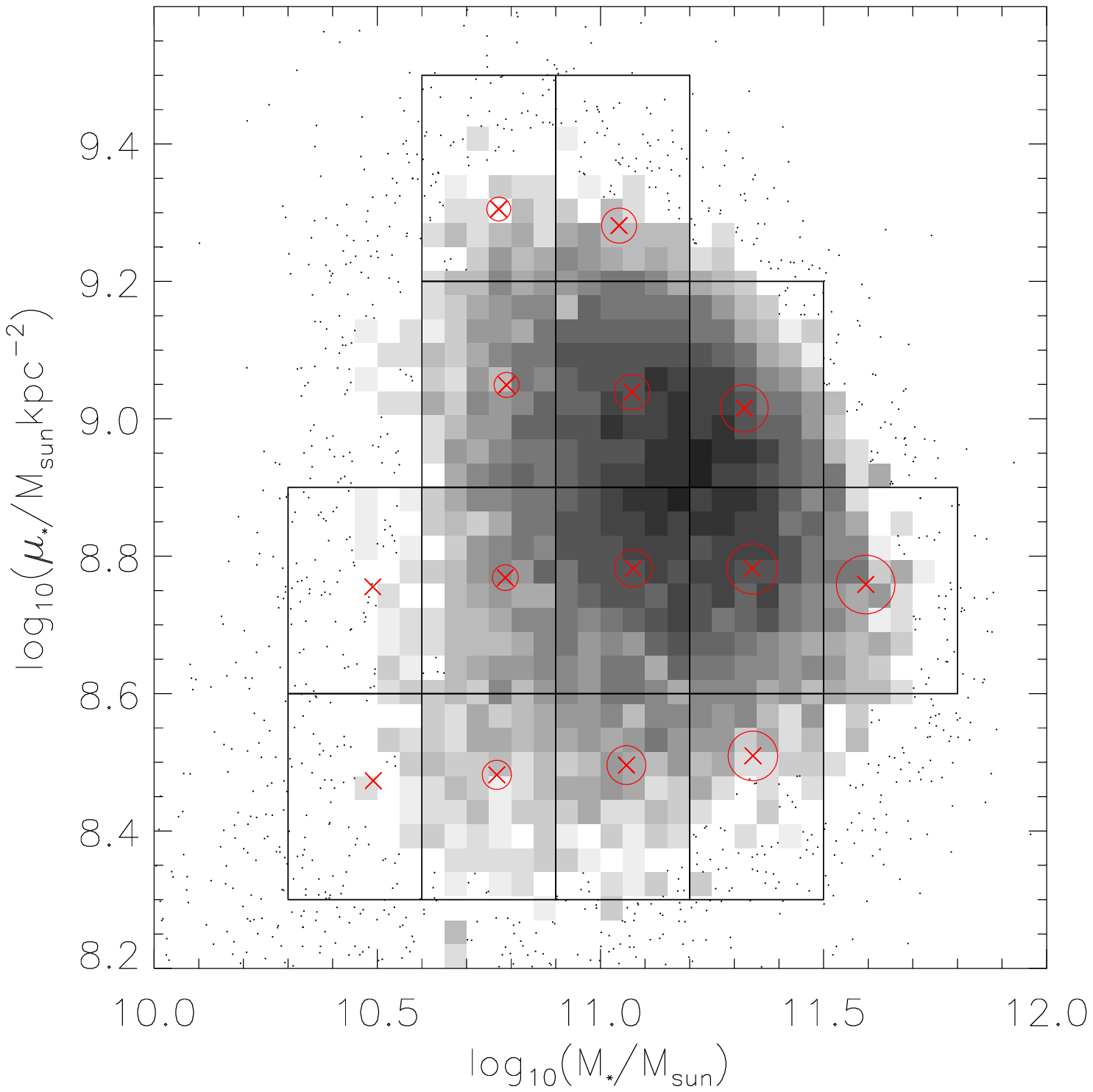,width=0.32\textwidth}
    \epsfig{figure=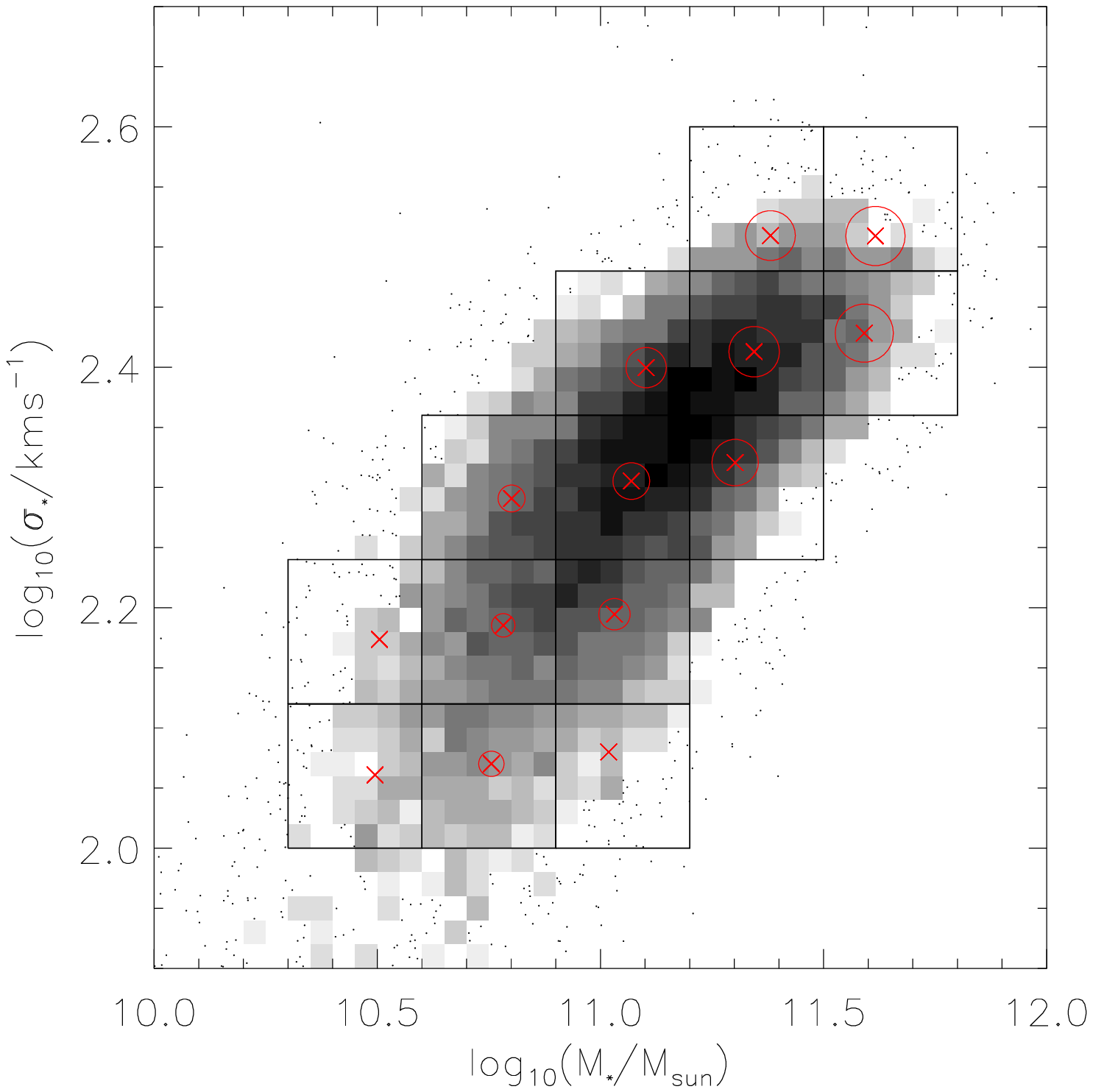,width=0.32\textwidth}
  \end{center}
  \caption{Distribution of the central galaxies in the SDSS/DR7 galaxy
    group catalog in the plane of (from left to right) stellar mass
    (\mstar) versus  optical color (\gr), \mstar\ versus surface
    stellar mass density (\mustar), and \mstar\ versus central stellar
    velocity dispersion (\vdisp).  Closed boxes indicate the
    subsamples analyzed in this paper, while the red cross in each box
    denotes the average of the sample galaxies. The red circle
    indicates the estimated halo mass (\mhalo), with the size of the
    circles being scaled by \lgmhalo.}
  \label{fig:cen_samples}
\end{figure*}

\begin{figure}
  \begin{center}
    \epsfig{figure=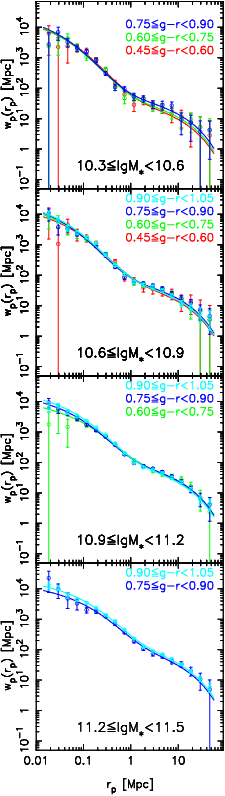, width=0.22\textwidth}
    \epsfig{figure=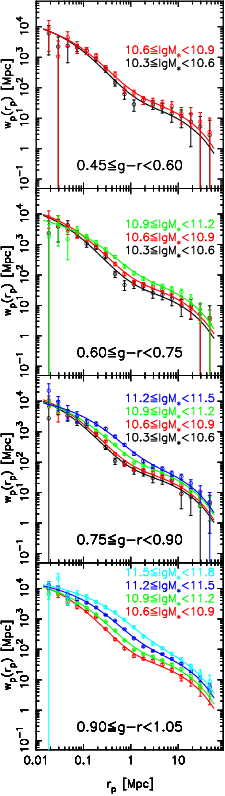, width=0.22\textwidth}
  \end{center}
  \caption{Projected cross-correlation function between subsamples of
    central galaxies and the reference galaxy sample, for central
    galaxies of different stellar mass (\mstar) and optical color
    (\gr), as indicated in each panel. Plotted in colored symbols with
    error bars are the results from the SDSS/DR7 galaxy group catalog,
    while the solid lines present the best-fit models (see the text
    for details).}
  \label{fig:wrp_mstar_gr}
\end{figure}

\begin{figure*}
  \begin{center}
    \epsfig{figure=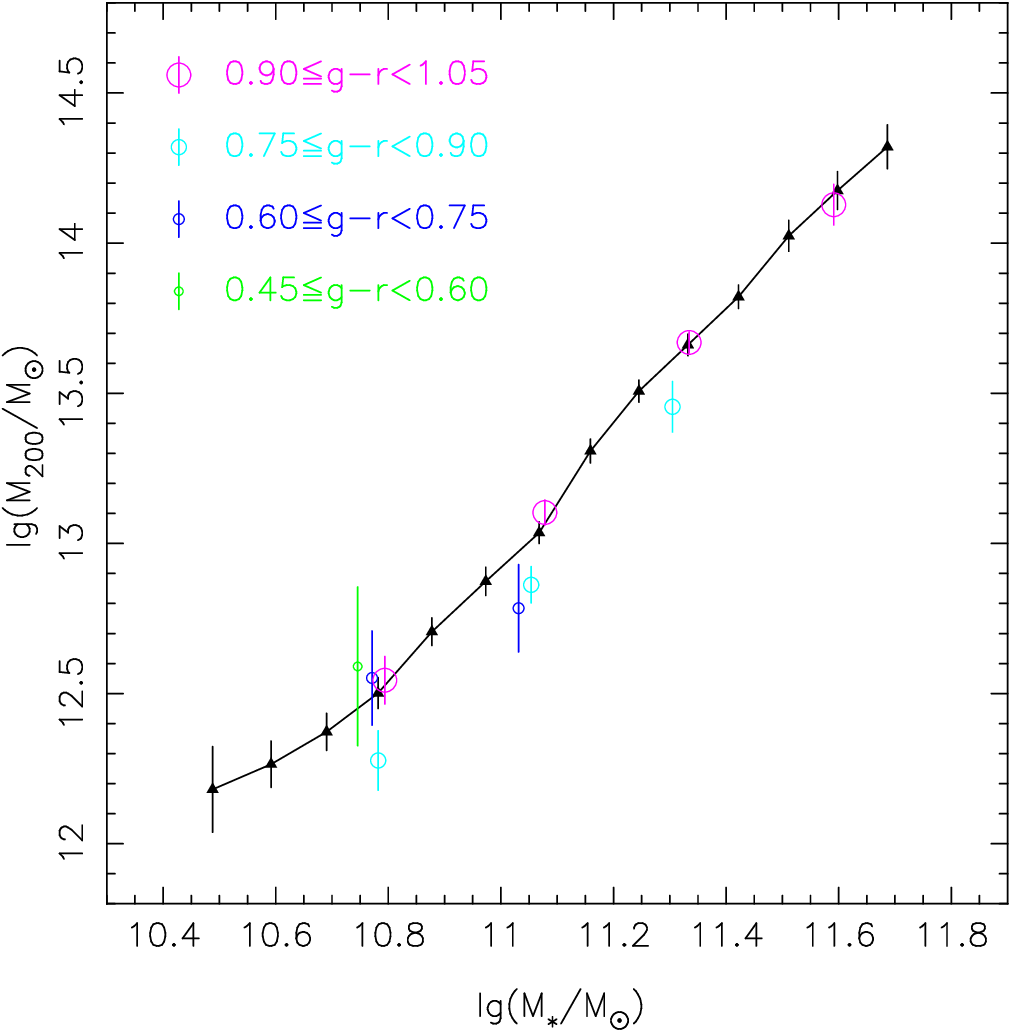, width=0.45\textwidth}
    \epsfig{figure=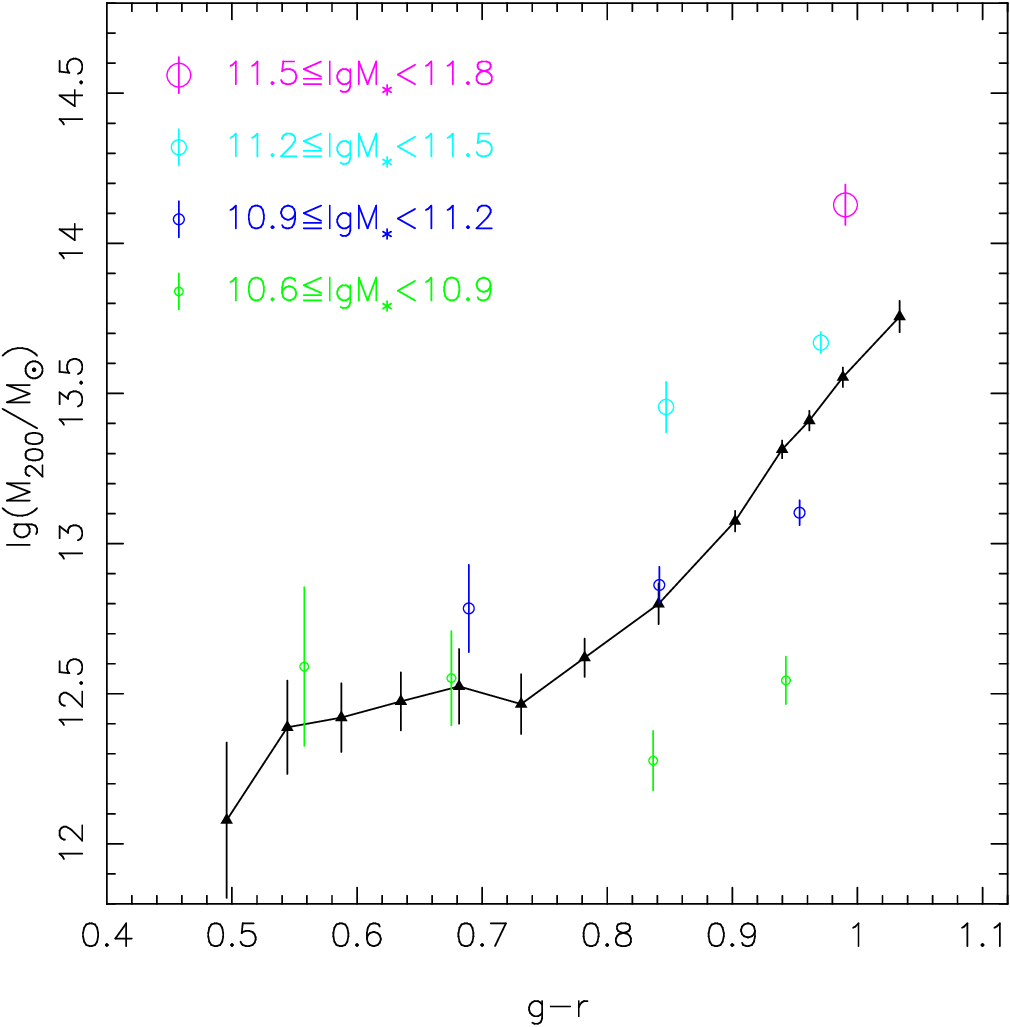, width=0.45\textwidth}
  \end{center}
  \caption{Dark matter halo mass as function of stellar mass (\mstar,
    left panel) and optical color (\gr, right panel), measured for the
    central galaxies of the SDSS/DR7 groups. Plotted in colored
    symbols are the results for the subsamples selected on the
    \mstar\ vs. \gr\ plane  as shown in Figure~\ref{fig:cen_samples},
    while the triangles connected with the solid line are for samples
    selected only by \mstar\ (left panel) or \gr\ (right panel).}
  \label{fig:mhalo_mstar_gr}
\end{figure*}

\subsection{Galaxy properties}

The galaxy properties needed for this work include stellar mass
(\mstar), surface stellar mass density (\mustar), optical color (\gr),
and central stellar velocity dispersion (\vdisp). We take the
measurements of these properties from both the NYU-VAGC and the
MPA/JHU SDSS database. The latter is publicly available at
http://www.mpa-garching.mpg.de/SDSS/DR7/, and is described in detail
in \citet{Brinchmann-04}. 

We use the optical color index defined by the $g$-band and $r$-band
Petrosian magnitudes, which are corrected for Galactic extinction, and
$K$-corrected to their values at $z=0.1$ using  the {\tt kcorrect}
code of \citet{Blanton-03b}.

A stellar mass accompanies the NYU-VAGC release for each galaxy in our
group catalog and reference sample. This was estimated by
\citet{Blanton-Roweis-07} based on the spectroscopically-measured
redshift of the galaxy and its ``Petrosian'' magnitude in the SDSS
five bands, assuming a universal initial mass function of
\citet{Chabrier-03} form and implicitly correcting for dust
attenuation. We have corrected the Petrosian mass to obtain a ``total
mass'' using the SDSS model magnitudes as described in Appendix A of
\citet{Guo-10}.  As demonstrated in \citet{Li-White-09}, the masses
from \citet{Blanton-Roweis-07} agree pretty well with those obtained
from the simple, single color estimator of \citet{Bell-03}, as well as
those derived by \citet{Kauffmann-03b} from a combination of SDSS
photometry and spectroscopy. In order to test whether a different
stellar mass definition will cause any bias in our analysis, we also
use the stellar masses from the MPA/JHU database obtained by
\citet{Kauffmann-03b}.

The surface stellar mass density of a galaxy is defined by
\mustar$=0.5$\mstar$/\pi R_{50,z}^2$, where $R_{50,z}$ is the radius
enclosing half of the total light in $z$-band, and \mstar\ is the
``total'' stellar mass from NYU-VAGC as described above. 

The SDSS spectroscopy provides a stellar velocity dispersion
($\sigma_\ast$) in the central 3$^{\prime\prime}$ of each galaxy. 
For disk-dominated galaxies, however, the $\sigma_\ast$ thus obtained
may be dominated by the global rotation rather than the random motion
of stars. Therefore we obtain the central stellar velocity dispersion 
for our galaxies based on both photometry and spectroscopy from SDSS, 
but in different ways for galaxies with and without a prominent bulge. 
We determine the significance of bulge based on data from \citet{Simard-11} 
who have carried out a careful bulge/disk decomposition for $\sim1.1$  
million galaxies using the SDSS imaging. A galaxy is classified to 
have a prominent bulge if the bulge-to-total luminosity ratio
$L_{bulge}/(L_{bulge}+L_{disk})>0.5$, or the effective radius of the
bulge $R_{e,bulge}>1.5^{\prime\prime}$. Otherwise, the galaxy is
classified to have no/weak bulge component. Here the bulge/disk
luminosities and radii are given by the $n=4$ bulge $+$ disk fitting
model performed to the SDSS $r$-band image (see \citealt{Simard-11}
for  detailed description of the model). According to the significance
of bulge and the signal-to-noise (S/N) of the SDSS spectrum, we 
divide the galaxies into three classes, and obtain \vdisp\ in different
ways for different classes.
\begin{itemize}
  \item For those galaxies with both a prominent bulge and a high S/N
    spectrum ($r$-band S/N$>10$), we directly use the
    \vdisp\ measurement provided in the MPA/JHU database. Following
    previous studies we have corrected the \vdisp\ of a galaxy to an
    aperture of its effective radius, adopting the relation found by
    \citet{Jorgensen-Franx-Kjaergaard-95}: $\sigma_{\ast,\rm
      corr}/\sigma_{\ast,\rm fib}=(8{\times}r_{\rm
      fib}/R_{50,r})^{0.04}$, where $r_{\rm fib}=1.5^{\prime\prime}$
    and $R_{50,r}$ is the radius enclosing half of the total light in
    $r$-band. At \vdisp$>100$kms$^{-1}$ where we perform this study,
    about 53\% of our galaxies fall into this class.
  \item For the galaxies with a prominent bulge but a low-quality
    spectrum  with S/N$\leq10$, we assign a \vdisp\ to each galaxy
    according to the  correlation between \vdisp\ and bulge luminosity
    $L_{\mbox{bulge}}$, which is given by the galaxies in the above
    class.  In practice, we bin all the galaxies from the above class
    into narrow intervals of $L_{\mbox{bulge}}$, and for a given
    interval we obtain the distribution function of \vdisp\ according
    to which we randomly generate a \vdisp\ for the galaxies falling
    in this interval. In this way both the median relation of
    \vdisp-$L_{\mbox{bulge}}$ in the high-S/N sample and the scatter
    of  the galaxies around the median relation are reproduced in the
    low-S/N sample. At \vdisp$>100$kms$^{-1}$ only a tiny fraction (1.4\%) of
    galaxies fall into this class. Our tests here show that dropping 
    this class would not impact our results in this paper. 
  \item For the galaxies without a prominent bulge, we follow previous
    studies \citep[e.g.][]{Sheth-03,Chae-10} to estimate \vdisp\ from
    converting the circular velocity $v_c$:
    $\sigma_\ast=v_{c}/\sqrt{2}$, assuming the singular isothermal
    sphere (SIS) model for the total mass distribution of a galaxy.
    The circular velocity is given by the Tully-Fisher relation:
    $\log_{10}v_c = (-0.135\pm0.006)(M_r-M_r^\ast)+(2.210\pm0.006)$
    with $M_r^\ast-5\log_{10}h=-20.332$ and a dispersion of
    $0.063\pm0.005$, which was found by \citet{Pizagno-07} for the
    SDSS $r$-band. At \vdisp$>100$kms$^{-1}$ a fraction of 45.6\%
    galaxies fall into this class.
\end{itemize}

We would like to point out that, although we have obtained the
central $\sigma_\ast$ differently for different morphologies, our results
to be presented below are essentially insensitive to the $\sigma_\ast$ 
measurements. We have repeated all the analyses using the $\sigma_\ast$
measured directly from the SDSS spectra, obtaining exactly the same 
conclusions. A part of these analyses is presented in Appendix of this
paper. 

\section{Methodology}
\label{sec:methods}

Detailed description of our methods for measuring galaxy clustering and 
dark matter halo masses can be found in our previous papers 
\citep{Li-06b, Li-06a, Li-12c}. In this section we briefly describe
these methods. Following these papers we estimate errors on all the 
measurements in the current work using the bootstrap resampling technique 
\citep{Barrow-Bhavsar-Sonoda-84}.

\subsection{Two-point cross-correlation function}

In this work we use two-point cross-correlation function (2PCCF) to 
quantify the clustering of our galaxies. For a given sample of galaxies
({\tt Sample Q}), this is measured with respect to the reference galaxy sample 
constructed above ({\tt Sample G}), over scales from a few tens of kpc up to 
a few tens of Mpc. On scales larger than a few Mpc, the amplitude of the 2PCCF 
provides a measure of the average mass for the dark matter halos that host the 
sampled galaxies.  First, the redshift-space 2PCCF, $\xi^{(s)}(r_p,\pi)$, between 
{\tt Sample Q} and {\tt G}, is estimated by
\begin{equation}
\xi^{(s)}(r_p,\pi) = \frac{N_{R}}{N_{G}}\frac{QG(r_{p},\pi)}{QR(r_{p},\pi)}-1.
\end{equation}
Here, $N_{R}$ and $ N_{G}$ are the number of galaxies in the random sample 
({\tt Sample R}) and in {\tt Sample G}; $r_{p}$ and $\pi$ are the pair separations
perpendicular and parallel to the line of sight;  $QG(r_{p},\pi)$ and  ${QR(r_{p},\pi)}$ 
are the counts of cross pairs between {\tt Sample Q} and  {\tt G}, and between
 {\tt Sample Q} and {\tt R}. The projected 2PCCF is then obtained by integrating 
$\xi^{(s)}(r_p,\pi)$ over $\pi$:
\begin{equation}
w_{p}(r_{p}) = \int_{-\infty}^{+\infty}\xi(r_{p},\pi)d\pi = \sum\xi(r_{p},\pi_{i})\Delta\pi_{i}. 
\end{equation}        
The summation for computing $w_p(r_p)$ runs from $\pi_1=-39.5h^{-1}$Mpc to
$\pi_{80}=39.5h^{-1}$Mpc, with $\Delta\pi_{i} = 1 h^{-1} Mpc$. We correct the 
effect of fiber collisions using the method detailed in \citet{Li-06a}.

\subsection{Dark matter halo mass of central galaxies}

For a given sample of central galaxies, we begin by estimating the 
redshift-space 2PCCF $\xi^{(s)}(r_p,\pi)$ and the projected 2PCCF
$w_{p}(r_{p})$ in the same way as described above. We then obtain 
the 2PCCF in real space, $\xi_{cg}(r)$, by the Abel transform of the 
$w_p(r_p)$. Next, the one-dimensional velocity dispersion profile (VDP) 
of satellite galaxies around the centrals is estimated by comparing 
$\xi^{(s)}(r_p,\pi)$ with $\xi_{cg}(r)$, through modeling the redshift-space 
distortion in $\xi^{(s)}(r_p,\pi)$. 

In order to estimate an average mass for the dark matter halos that
host these central galaxies, we have calibrated the relationship between 
the one-dimensional velocity dispersion and the dark matter halo mass,
by applying the same method to $N$-body cosmological simulations.
Finally, based on this relationship, we determine an average halo mass 
for the sample of central galaxies from their one-dimensional 
velocity dispersion.

It was demonstrated in L12 that the halo mass thus estimated for central 
galaxies of given luminosity 
agrees very well with that obtained by \citet{Mandelbaum-06} by stacking
the weak lensing signals of SDSS galaxies.
We have also compared the halo mass versus stellar mass relation 
obtained from our method with that in \citet{Zu-Mandelbaum-15} and
\citet{Mandelbaum-15}, which were measured from galaxy-galaxy lensing but 
had smaller statistical errors and better mass modelling compared to
\citet{Mandelbaum-06}. Detailed description of our
methodology, as well as comparisons with previous work and tests on
the calibration, can be found in L12. 

\section{Results}
\label{sec:results}

\subsection{Clustering and halo mass of central galaxies}
\label{sec:centrals}

We begin with the central galaxies in our group catalog,   estimating
their redshift-space clustering and dark matter halo mass as function
of  their physical properties including stellar mass (\mstar), optical
color (\gr), surface stellar mass density (\mustar) and central
stellar velocity dispersion (\vdisp). To this end, we have selected a
number of subsamples on the plane of \mstar\ versus each of the other
parameters: \gr, \mustar\ and \vdisp. The selection is shown in
Figure~\ref{fig:cen_samples}, which displays all the  central galaxies
with small dots in each of the three planes, with the subsamples in
each plane enclosed by the black boxes. The  average values of the two
parameters for the selection is marked with a cross for each
subsample. This selection scheme gives us at least two intervals in
one parameter when the other is limited to a given interval,  thus
allowing us to study the dependence of clustering and halo mass on one
of the two parameters while fixing the other.

Figure~\ref{fig:wrp_mstar_gr} presents the measurements of \wrp\ for
the subsamples of central galaxies selected on the
\mstar\ vs. \gr\ plane. We plot the results for different
\gr\ subsamples but fixed \mstar\ in the left-hand panels, and the
results for different \mstar\ subsamples but fixed \gr\ in the
right-hand panels. At fixed stellar mass and given projected
separation $r_p$, \wrp\ shows very weak or no dependence on \gr, and
this result holds for all the masses ($10.3<$\lgmstar$<11.5$) and
separations ($\sim20$kpc$<r_p<\sim50$Mpc) probed here. In contrast,
the \wrp\ at fixed \gr\ presents strong  dependence on stellar mass,
in terms of both amplitude and slope.  This effect is true for all the
\gr\ bins except the bluest color bin  ($0.45\le$\gr$\le0.60$) where
we have only two subsamples due to  the relatively narrow range in
stellar mass. At given \gr\ and on scales between $\sim100$kpc and a
few Mpc, the \wrp\ amplitude increases, while its slope flattens with
increasing \mstar. As a result, the transition between the
``one-halo'' and ``two-halo'' regimes occur at larger radii for higher
masses, indicating that high-mass centrals are hosted by dark matter
halos with larger virial radii (thus higher dark matter masses) when
compared to low-mass centrals of the same \gr\ color. 

\begin{figure}
  \begin{center}
    \epsfig{figure=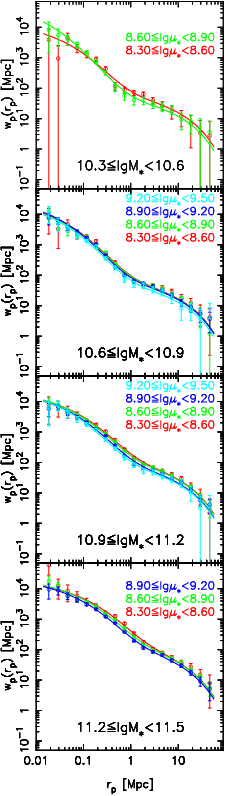,width=0.22\textwidth}
    \epsfig{figure=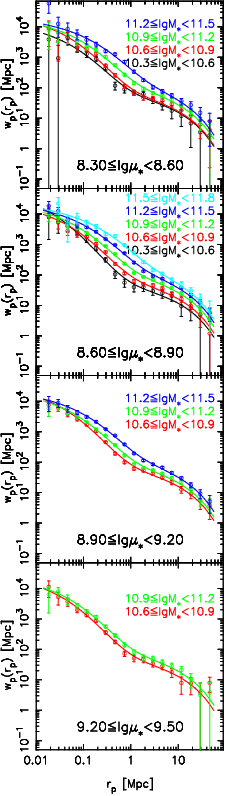,width=0.22\textwidth}
  \end{center}
  \caption{Projected cross-correlation function between subsamples of
    central galaxies and the reference galaxy sample, for central
    galaxies of different stellar mass (\mstar) and surface stellar
    mass density (\mustar), as indicated in each panel. Plotted in
    colored symbols with error bars are the results from the SDSS/DR7
    galaxy group catalog, while the solid lines present the best-fit
    models (see the text for details).}
  \label{fig:wrp_mstar_mustar}
\end{figure}
\begin{figure*}
  \begin{center}
    \epsfig{figure=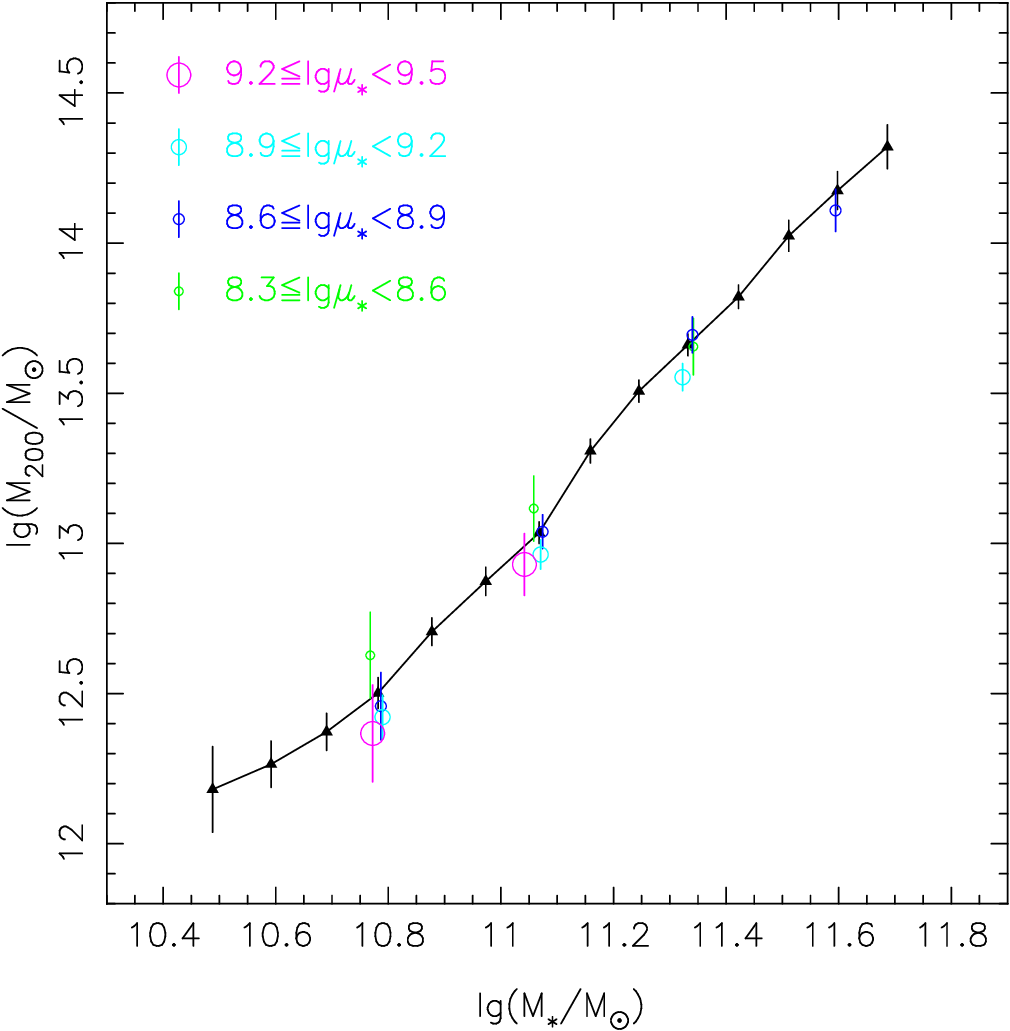,width=0.45\textwidth}
    \epsfig{figure=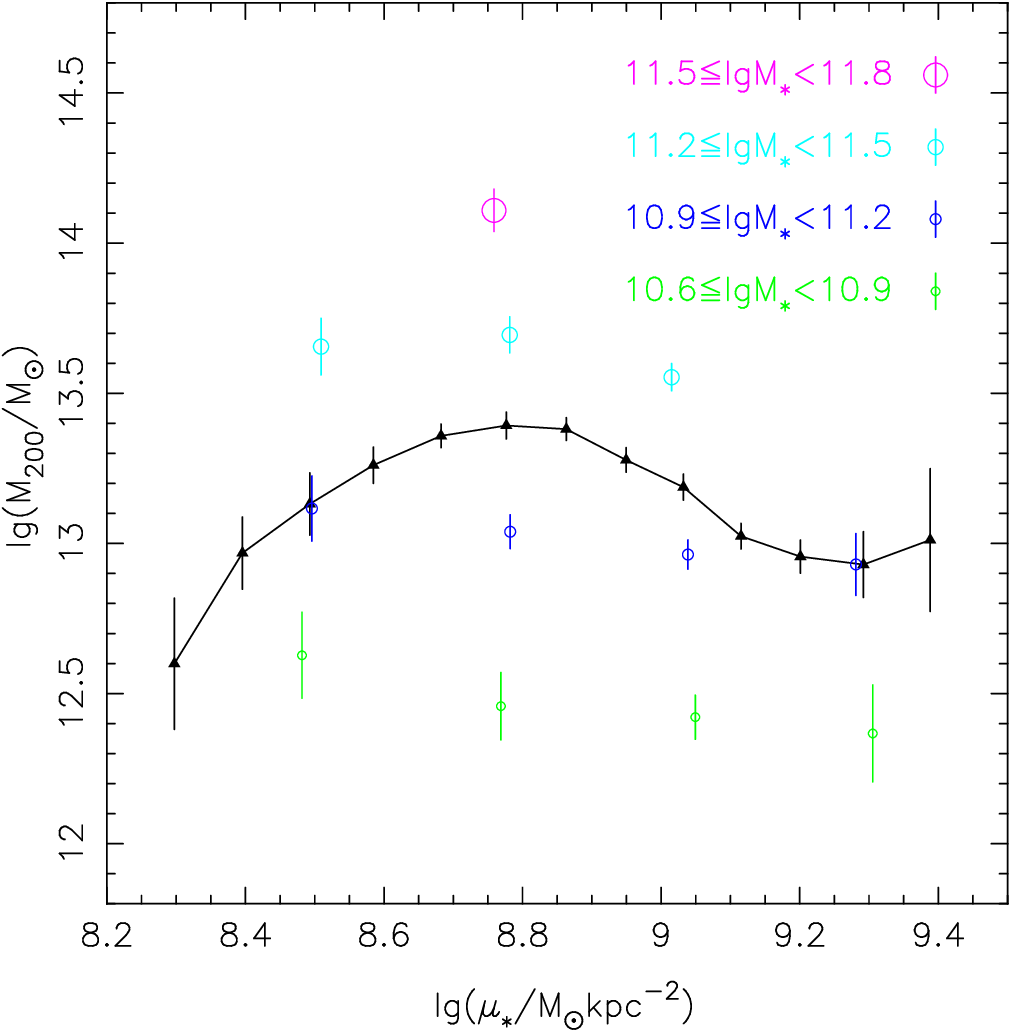,width=0.45\textwidth}
  \end{center}
  \caption{Dark matter halo mass as function of stellar mass (\mstar,
    left panel) and surface stellar mass density (\mustar, right
    panel), measured for the central galaxies of the SDSS/DR7
    groups. Plotted in colored symbols are the results for the
    subsamples selected on the \mstar\ vs. \mustar\ plane  as shown in
    Figure~\ref{fig:cen_samples}, while the triangles connected with
    the solid line are for samples selected only by \mstar\ (left
    panel) or \mustar\ (right panel).}
  \label{fig:mhalo_mstar_mustar}
\end{figure*}

We present the correlation of halo mass with stellar mass and color in
Figure~\ref{fig:mhalo_mstar_gr}, where the halo mass (\mhalo)
estimated for all the central galaxy subsamples is plotted against
both \mstar\ (left panel) and \gr\ (right panel). For comparison we
have done the same analysis for a set of subsamples selected by
\mstar\ without further dividing the galaxies in each subsample into
intervals of \gr, as well as a set of \gr\ subsamples without further
dividing the galaxies in each subsample into \mstar\ intervals. This
analysis leads to an {\em average} relation between \mhalo\ and
\mstar\ (left panel), and between \mhalo\ and \gr\ (right  panel),
which are plotted as triangles connected with a solid line in the
figure.  The figure shows that \mhalo\ is correlated with both
\mstar\ and \gr, but the correlation with \mstar\ is apparently much
stronger than the correlation with \gr, in terms of both the scatter
and slope of the relations.  The rms scatter of the different
\gr\ subsamples around the average relation between \mhalo\ and
\mstar\ is 26.1\%, compared to  138\% for the different
\mstar\ subsamples around the \mhalo--(\gr) relation.  The average
relation between \mhalo\ and \mstar\ can be roughly described  by a
straight line with a steep slope.  In contrast, the average relation
between \mhalo\ and \gr\ shows a similarly steep (but more scattering)
relation only  for red galaxies with (\gr)$\ga 0.7$. For central
galaxies of bluer colors, the average halo mass is nearly constant at
$\sim3\times10^{12}$M$_\odot$, close to the halo mass of the Milky
Way.  We note that, in the stellar mass range of
$10.6\le$\lgmstar$<10.9$, the halo mass appears to even show a
non-monotonic dependence on color (see the green symbols in the
right-hand panel), with both blue and red galaxies being hosted by dark
halos of mass \lgmhalo$\approx12.5$ and the galaxies with intermediate
color ($0.75\le$\gr$<0.9$) moving in slightly lighter halos
(\lgmhalo$\sim12.3$). 

Figure~\ref{fig:wrp_mstar_mustar} and~\ref{fig:mhalo_mstar_mustar}
present the \wrp\ and \mhalo\ estimates for subsamples selected on the
plane of \mstar\ and \mustar, in the same format and with the same
symbols as above. The results are quite similar to what are found from
the previous two figures, in the sense that \wrp\ and \mhalo\ at fixed
\mustar\ show strong dependence on stellar mass,  with little residual
correlation with \mustar\ when \mstar\ is fixed. The halo mass
averaged  over all stellar masses depends on \mustar\ in a complicated
way:  \mhalo\ increases with increasing \mustar, reaching a maximum of
\lgmhalo~$\approx13.4$ at $8.6\le$\lgmustar$<8.9$, before it declines
and becomes constant at \lgmustar$\ga9.2$. At given \mustar, the
different \mstar\ subsamples scatter by an order of 1-1.5 magnitude in
halo mass, with an overall rms scatter of 155\% around the average
relation between \mhalo\ and \mstar. For comparison, the rms scatter
is only 19.7\% for the \mustar\ subsamples around the
\mhalo--\mustar\ relation.

\begin{figure}
  \begin{center}
    \epsfig{figure=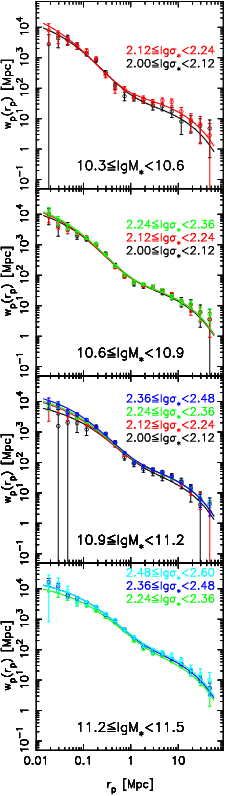,width=0.22\textwidth}
    \epsfig{figure=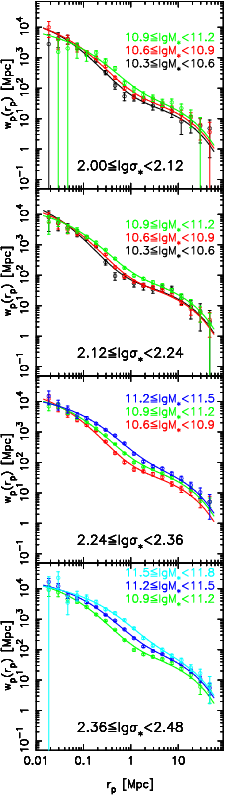,width=0.22\textwidth}
  \end{center}
  \caption{Projected cross-correlation function between subsamples of
    central galaxies and the reference galaxy sample, for central
    galaxies of different stellar mass (\mstar) and central stellar
    velocity dispersion (\vdisp), as indicated in each panel. Plotted
    in colored symbols with error bars are the results from the
    SDSS/DR7 galaxy group catalog, while the solid lines present the
    best-fit models (see the text for details).}
  \label{fig:wrp_mstar_vdisp}
\end{figure}

\begin{figure*}
  \begin{center}
    \epsfig{figure=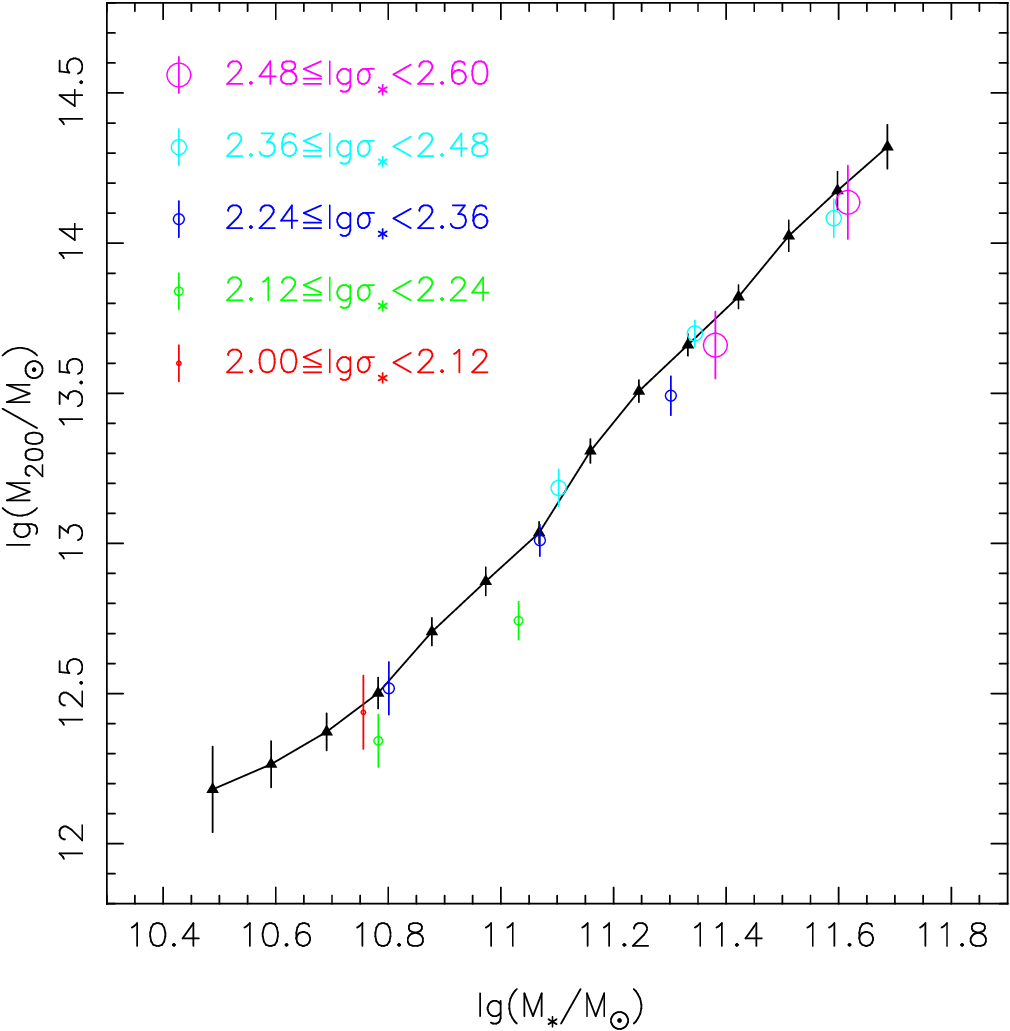,width=0.45\textwidth}
    \epsfig{figure=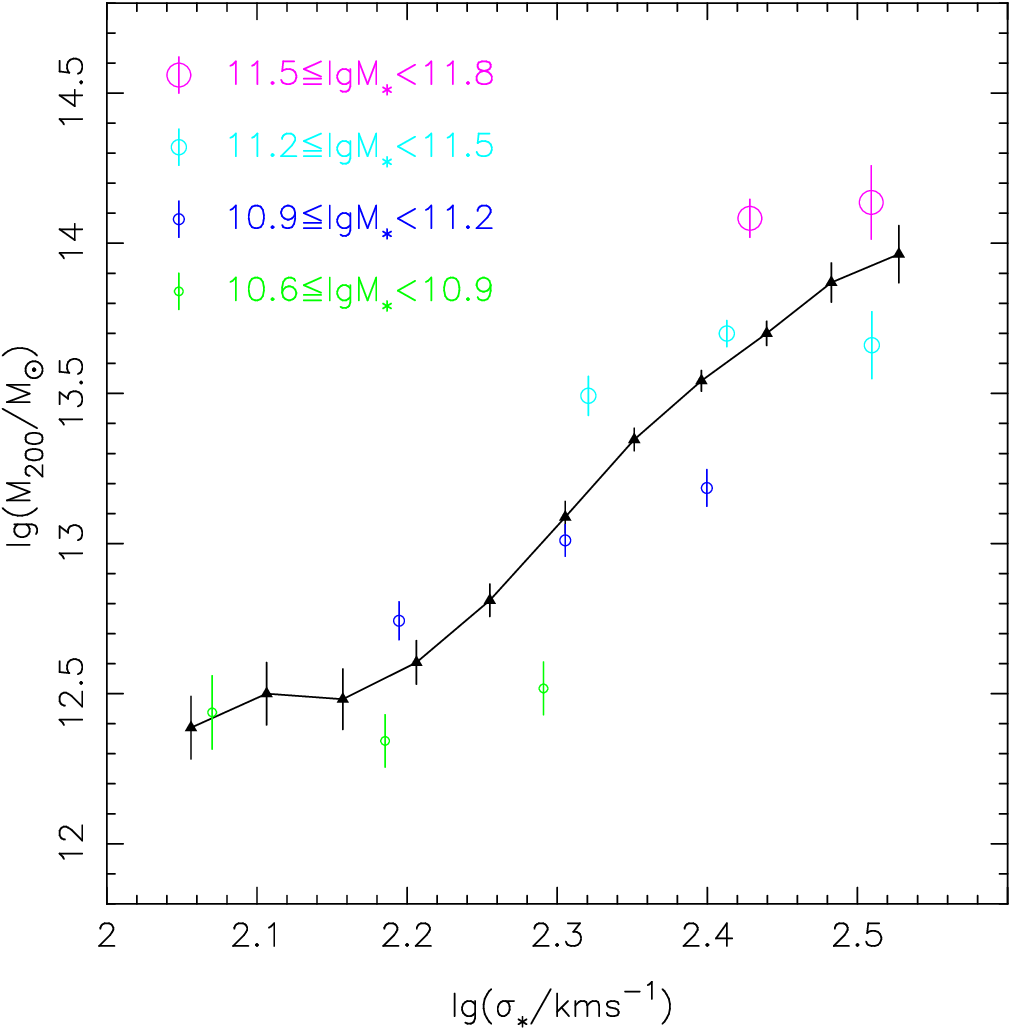,width=0.45\textwidth}
  \end{center}
  \caption{Dark matter halo mass as function of stellar mass (\mstar,
    left panel) and central stellar velocity dispersion (\vdisp, right
    panel), measured for the central galaxies of the SDSS/DR7
    groups. Plotted in colored symbols are the results for the
    subsamples selected on the \mstar\ vs. \vdisp\ plane  as shown in
    Figure~\ref{fig:cen_samples}, while the triangles connected with
    the solid line are for samples selected only by \mstar\ (left
    panel) or \vdisp\ (right panel).}
  \label{fig:mhalo_mstar_vdisp}
\end{figure*}

Figure~\ref{fig:wrp_mstar_vdisp} and~\ref{fig:mhalo_mstar_vdisp}
present the measurements of $w_p(r_p)$ and halo mass for the
subsamples selected on the \mstar--\vdisp\ plane.  These results were
presented in Figures 2 and 4 in L13, and are repeated here for
completeness and comparison. Similarly, \mhalo\ varies little  with
\vdisp\ at fixed \mstar, but shows strong and tight correlation with
\mstar\ when \vdisp\ is limited to a narrow range. The rms scatter is
15.2\% (73.2\%) for the subsamples of \vdisp\ (\mstar) around the
average relation of \mhalo\ with \mstar\ (\mustar). 


Figures~\ref{fig:wrp_mstar_gr} -\ref{fig:mhalo_mstar_vdisp}
echo the well-established tight correlation between the stellar mass
and halo mass for central galaxies in dark matter halos. More
importantly, these figures clearly show that the halo mass--stellar
mass correlation holds tight  even when \vdisp, \gr\ and \mustar\ are
limited to a narrow range. We conclude that, among the physical
parameters considered and for central galaxies with \lgmstar~$>10.3$,
stellar mass is the galaxy property that is  most closely correlated
with dark matter halo mass. Both \vdisp\ and \gr\  are correlated with
halo mass, with the former showing a steeper slope and a smaller
scatter, but these correlations are largely driven by the correlation
between halo mass and stellar mass. The \mustar\  is not correlated
with halo mass for the central galaxies as a whole.  At fixed stellar
mass, the estimated halo mass shows very little correlation with
\mustar, although we note that the halo mass appears to slightly
decline with increasing \mustar, particularly in the lowest stellar
mass bin ($10.6\le$\lgmstar$<10.9$).  The lack of correlation between
\mustar\ and halo mass at fixed stellar mass implies that the
mass--size relation of central galaxies, as  well established by the
large samples of both high-z and low-z galaxies, depends little on
their host halo mass. 

We summarize the co-dependence of halo mass on these physical parameters
by indicating the estimated halo mass for each sample with a red
circle  in Figure~\ref{fig:cen_samples}, where the size of the circle
is scaled linearly with \lgmhalo. The strong correlation of halo mass
with stellar mass, as well as the weak correlation with other
parameters, can be visually identified in this figure.
We conclude that, for central galaxies, the correlation of halo mass 
with stellar mass is still the tightest compared to the correlation 
with other properties.

\begin{figure}
  \begin{center}
    \epsfig{figure=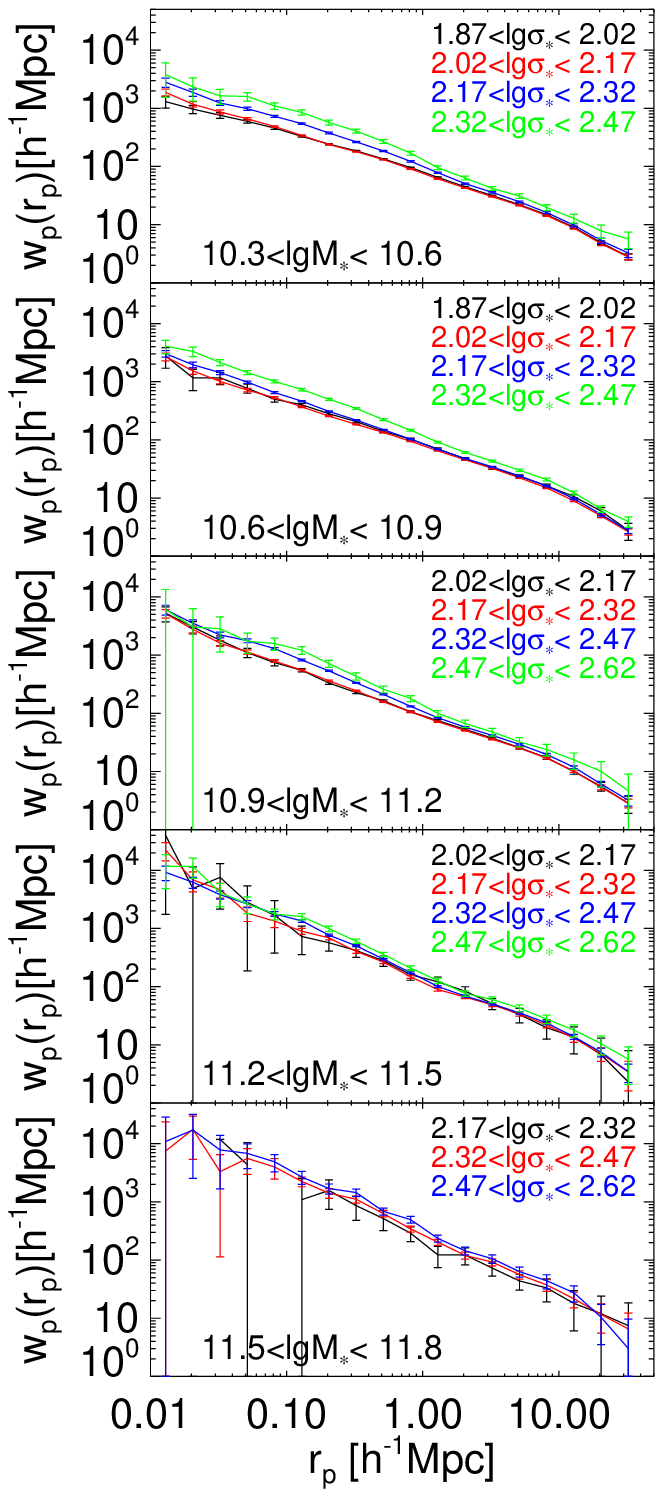,width=0.23\textwidth,height=0.6\textwidth}
    \epsfig{figure=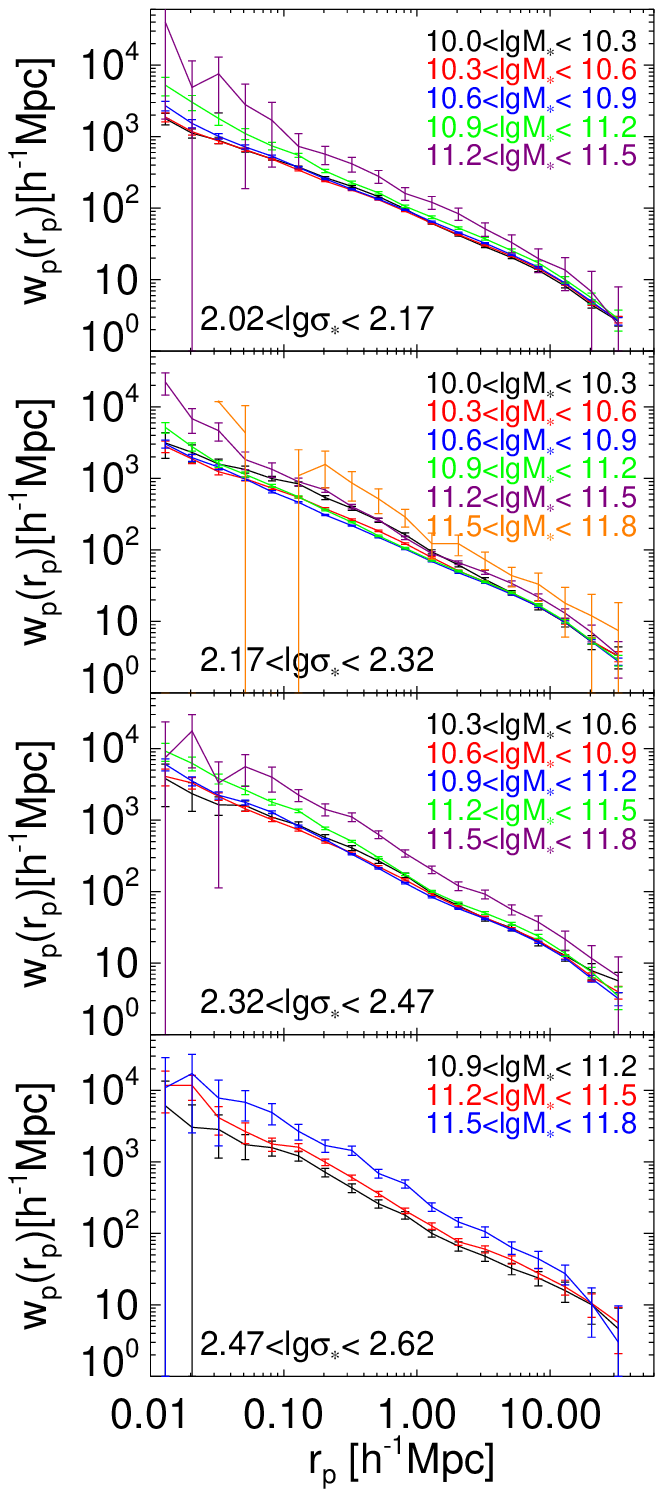,width=0.23\textwidth,height=0.6\textwidth}    
  \end{center}
  \caption{Projected cross-correlation function \wrp\ of subsamples of
    galaxies selected  by \mstar\ and \vdisp, with respect to the
    reference galaxy sample. Left panels show \wrp\ measurements in
    different intervals of \vdisp\ but fixed \mstar.  Right panels show
    the measurements for different \mstar\ intervals but fixed
    \vdisp.}
  \label{fig:all_wrp_mstar_vdisp}
\end{figure}

%%%%%%%%%%%%%%%%%%%%%%%%%%%%%%%%%%%%%%%%%%%%%%%%%%%%%%%%%%%%%
\subsection{Clustering of the full galaxy population and satellite galaxies}
\label{sec:all}

In this section we consider the full galaxy sample including both
central and satellite galaxies. Since our methodology of estimating
halo masses is applicable only for central galaxies,  in what follows
we will focus on measuring \wrp\ for subsamples selected from the full
sample, examining its co-dependence on the same set of  physical
properties analyzed above. Following the previous section, we select
subsamples of galaxies on two-dimensional planes formed by \mstar\ and
one of the other properties including \gr, \mustar\ and \vdisp, and
for each subsample we then estimate the project CCF \wrp\ with respect
to the reference galaxy sample using the same method. 

We start by measuring the \wrp\ for subsamples selected on the
\mstar--\vdisp\ plane.  Figure~\ref{fig:all_wrp_mstar_vdisp} presents
in the left panels the \wrp\ measurements at fixed \mstar\ but in
different \vdisp\ intervals, and in the right panels the measurements
at fixed \vdisp\ but different \mstar. The ranges of \mstar\ and
\vdisp\ used to select the subsamples are indicated in each panel.  An
overall impression from the figure is that, in almost all the panels
we see systematic shift in \wrp\ with increasing \mstar\ (or \vdisp)
when \vdisp\ (or \mstar) is fixed. At fixed \vdisp, the mass
dependence of the clustering power is mainly seen at the highest
stellar masses with \lgmstar$\ga 11$.  It is not surprising to note
that such mass dependence is most pronounced in the highest
\vdisp\ bin (the bottom-right panel), which is expected given the
known correlation between \mstar\ and \vdisp.  At fixed \mstar, the
\wrp\ increases with increasing \vdisp, and similarly the effect
appears to be stronger at higher \vdisp.

\begin{figure*}
  \begin{center}
    \epsfig{figure=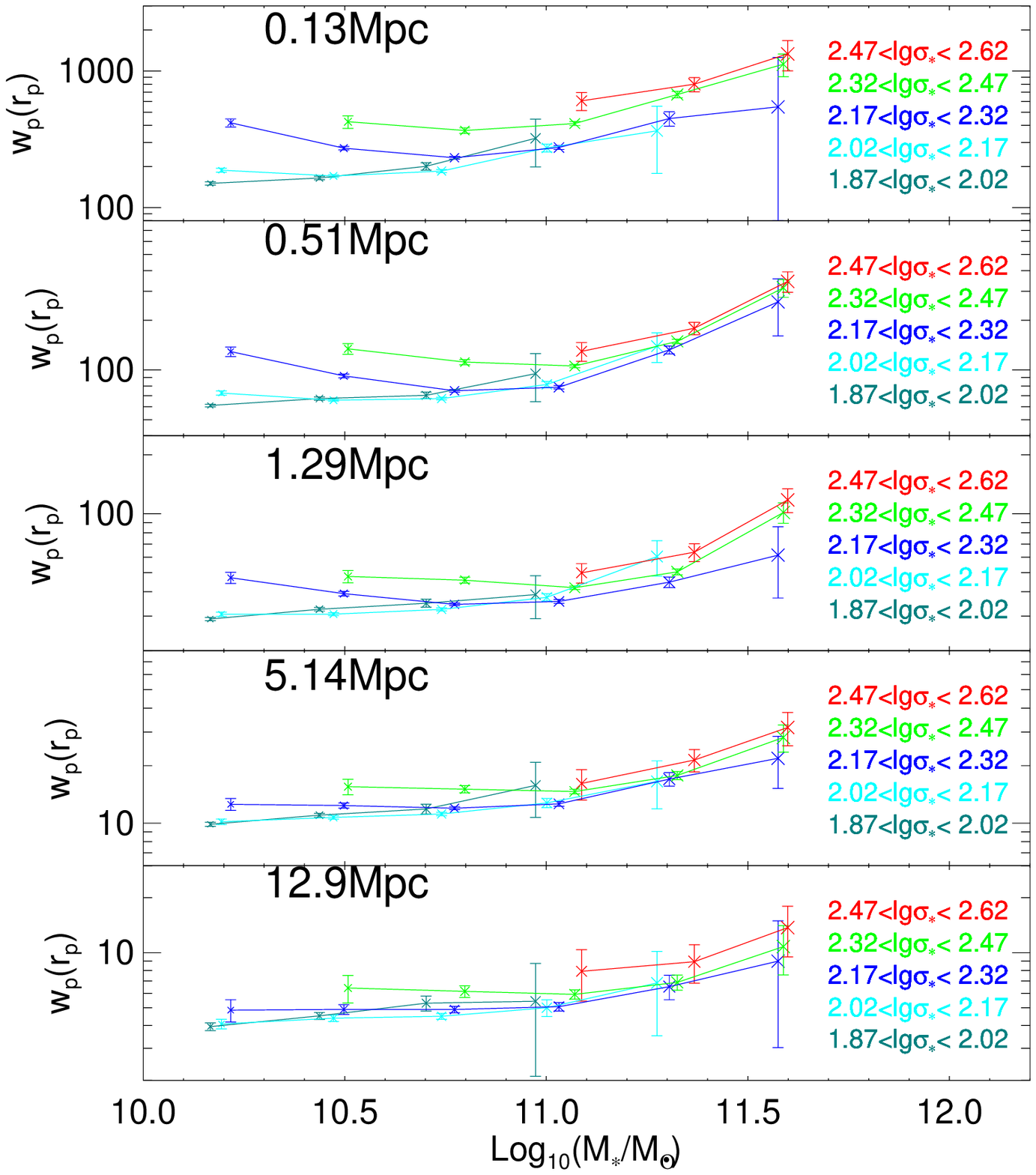,width=0.45\textwidth}
    \epsfig{figure=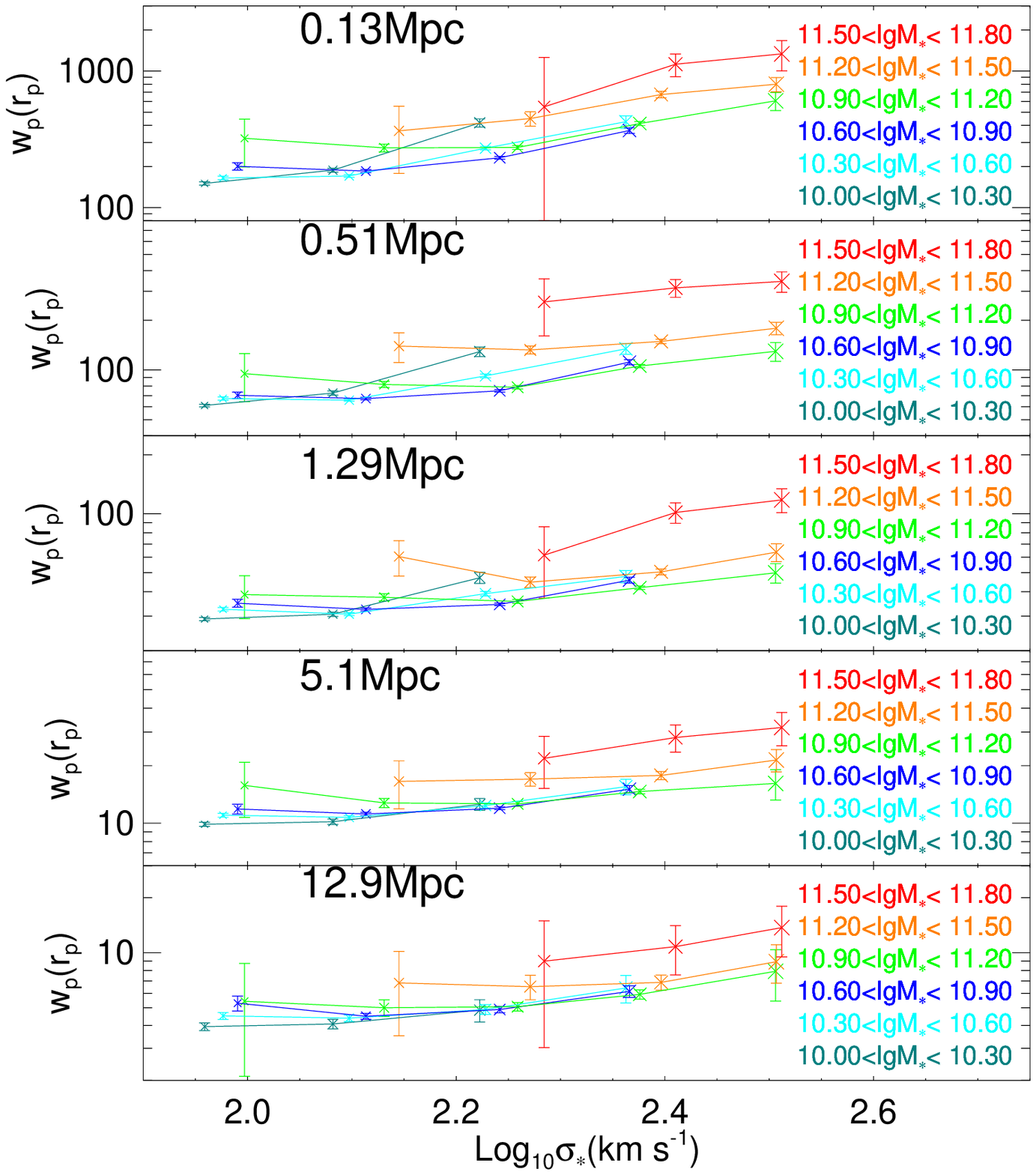,width=0.45\textwidth}
  \end{center}
  \caption{Projected cross-correlation function \wrp\ measured at
    fixed scale as indicated in each panel, as a function of
    \mstar\ (left panels) and \vdisp\ (right panels). In the left
    panels, the colored symbols/lines are for different
    \vdisp\ intervals but fixed \mstar, and those in the right panels
    are for different \mstar\ intervals but fixed \vdisp.}
  \label{fig:all_bias_mstar_vdisp}
\end{figure*}

These trends are more clearly seen in
Figure~\ref{fig:all_bias_mstar_vdisp} where we plot \wrp\ measured at
given projected separation $r_p$ as a function of \mstar\ for
subsamples of different \vdisp\ (left panels), and as a function of 
\vdisp\ for subsamples of different \mstar\ (right panels). In the left panels,
both the weak mass dependence at low masses and the strong mass
dependence  at the massive end are apparently seen. On average the
amplitude of \wrp\ changes little when one goes from the lowest masses
probed here (\mstar$\ga 10^{10}$M$_\odot$) up to \mstar$\sim
10^{11}$M$_\odot$,  before it increases remarkably at higher masses by
at most a factor 2-3 depending on scale and \vdisp. This result is
consistent with the well-established behavior of the relative bias
factor ($b/b_\ast$) of galaxy clustering measured in SDSS and other
surveys, which depends weakly on stellar
mass for galaxies with \mstar\ below the characteristic mass of the
stellar mass function but quickly increases at higher masses
\citep[e.g.][]{Tegmark-04, Norberg-01, Zehavi-04, Li-06b}. It is
interesting to see that, at high masses with \mstar$\ga
10^{11}$M$_\odot$ and at intermediate to large scales with $r_{p} > 0.5 Mpc$, 
the \wrp\ at fixed \mstar\ shows little dependence on \vdisp. 
In contrast, at lower masses and at all scales probed, the \wrp\ at fixed
\mstar\ appears to depend on \vdisp, with the effect being more marked
at smaller $r_p$. 

The \wrp\ as a function of \vdisp, plotted in the right-hand panels of
the figure, presents similar behaviors to what we found in the
left-hand panels, in the sense that the overall amplitude of \wrp\ at
given $r_p$ increases with increasing \vdisp, by a factor of 2-3 on
average.  However, when further dividing the galaxies by stellar mass,
we find that the increase of \wrp\ at high-\vdisp\ ($\ga 200$
kms$^{-1}$) is actually driven by the increase of \mstar\ with
\vdisp. In this  \vdisp\ regime, the \wrp\ shows little dependence on
\vdisp, while increasing dramatically with increasing \mstar\ at given
\vdisp. At \vdisp\ smaller than $\sim200$ kms$^{-1}$, in contrast, it
is the reverse that is in fact the case: the increase of \wrp\ with
\vdisp\ seems to be driven by \vdisp\ itself, an effect that is almost
independent of \mstar. 

\begin{figure*}
  \begin{center}
    \epsfig{figure=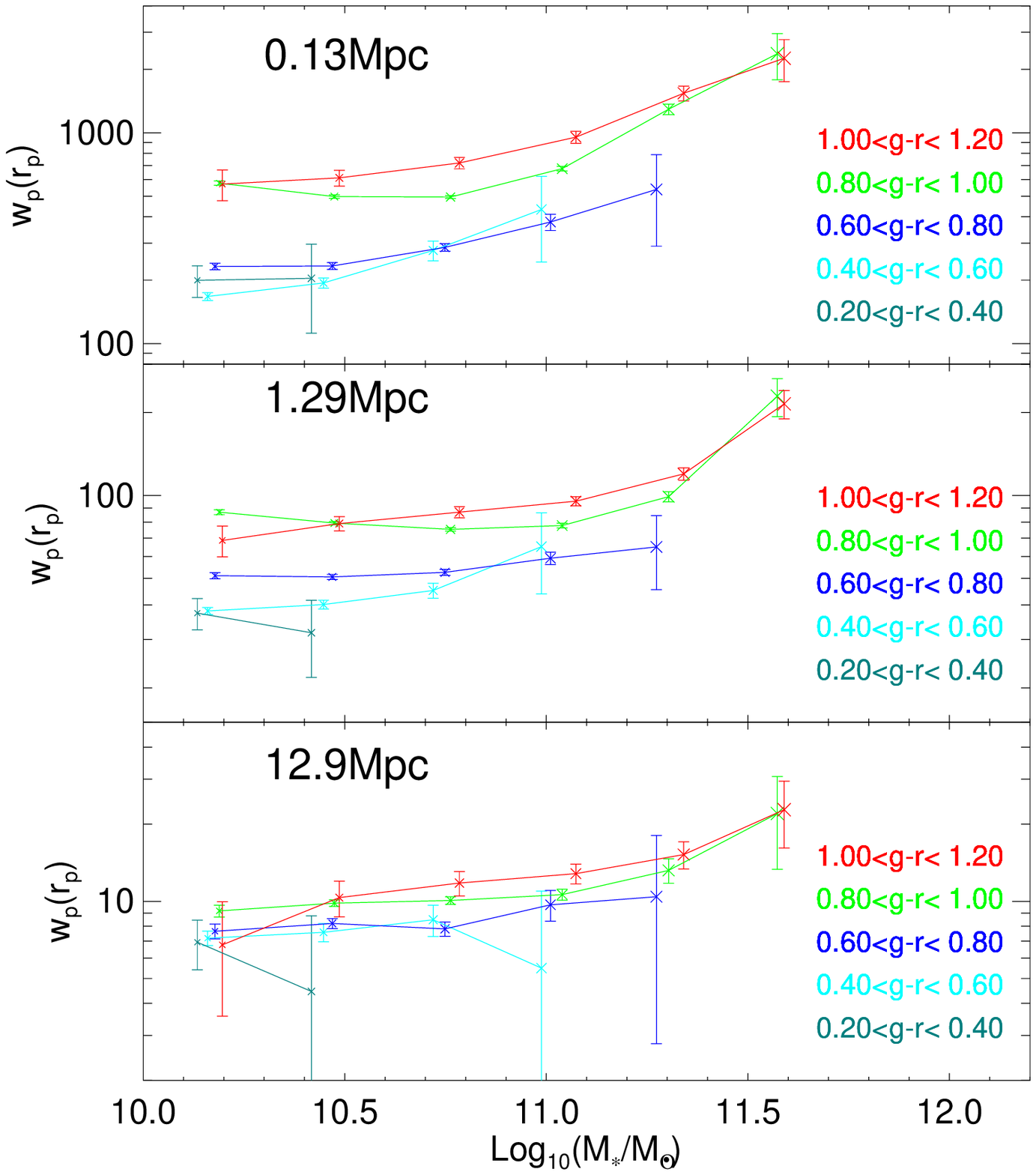,width=0.45\textwidth}
    \epsfig{figure=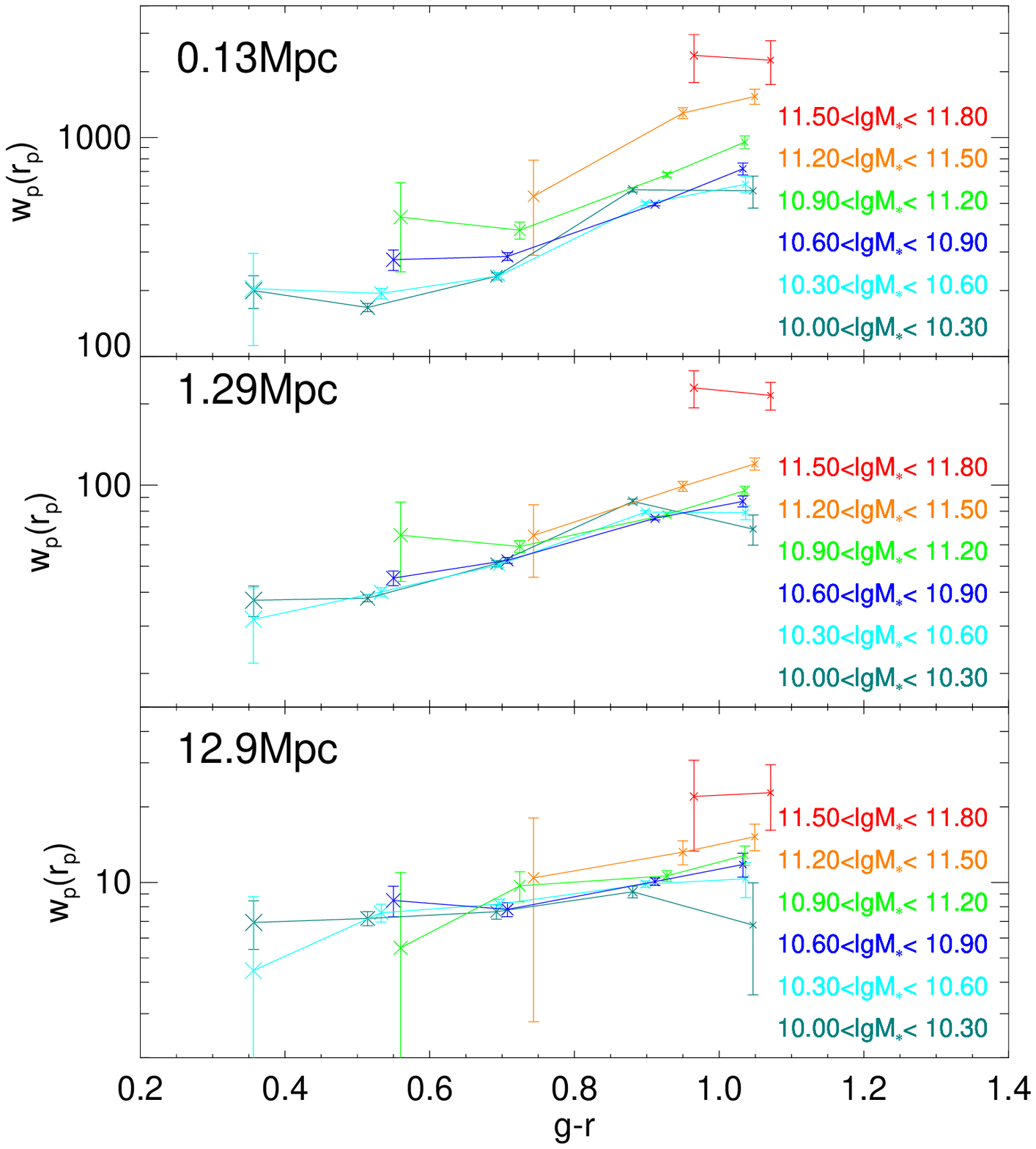,width=0.45\textwidth}
  \end{center}
  \caption{Projected cross-correlation function \wrp\ measured at
    fixed  scale as indicated in each panel, as a function of
    \mstar\ for subsamples of different \gr\ (left panels) and as a
    function of \gr\ for subsamples of different \mstar\ (right
    panels).  }
  \label{fig:all_bias_mstar_color}
\end{figure*}

We now extend the analysis by examining the co-dependence of  \wrp\ on
stellar mass and optical color (\gr). The \wrp\ measurements for the
subsamples selected by the two properties are presented in
Figure~\ref{fig:all_bias_mstar_color}.  Overall the results are
similar to what we see above for the  subsamples selected by
\mstar\ and \vdisp. At high masses ($>$ a few $\times
10^{11}$M$_\odot$) and red colors (\gr$\ga 0.8$), the clustering
amplitude depends strongly on \mstar\ and weakly on \gr. Conversely,
at lower masses and bluer colors, there is very weak or no dependence
on stellar mass, and the dependence on  color is clearly seen. These
results are also broadly consistent with previous studies which
measured the auto-correlation function of galaxies as a function of
both stellar mass and color \citep[e.g.][]{Li-06b}. 

\begin{figure*}[t]
  \begin{center}
    \epsfig{figure=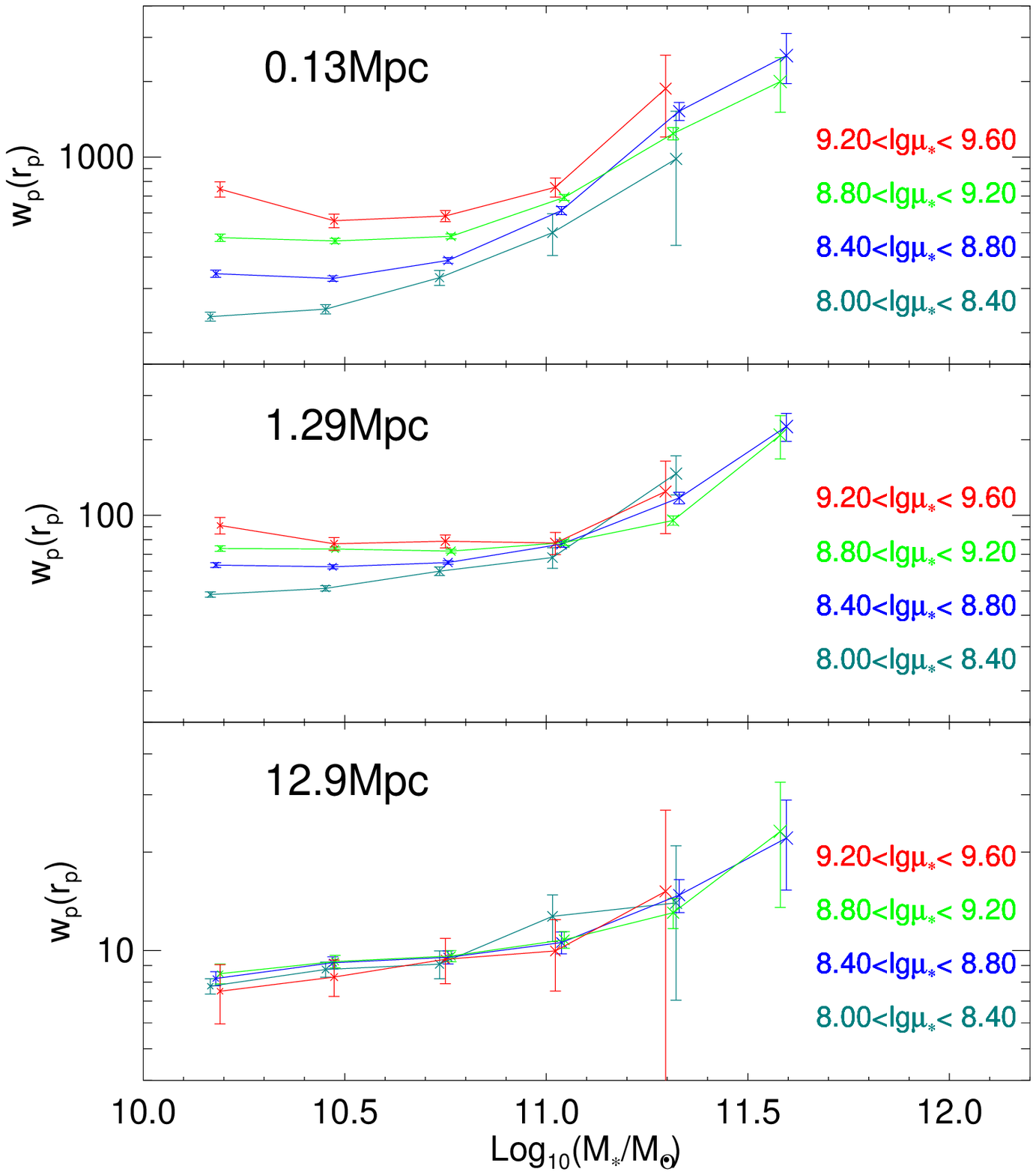,width=0.48\textwidth}
    \epsfig{figure=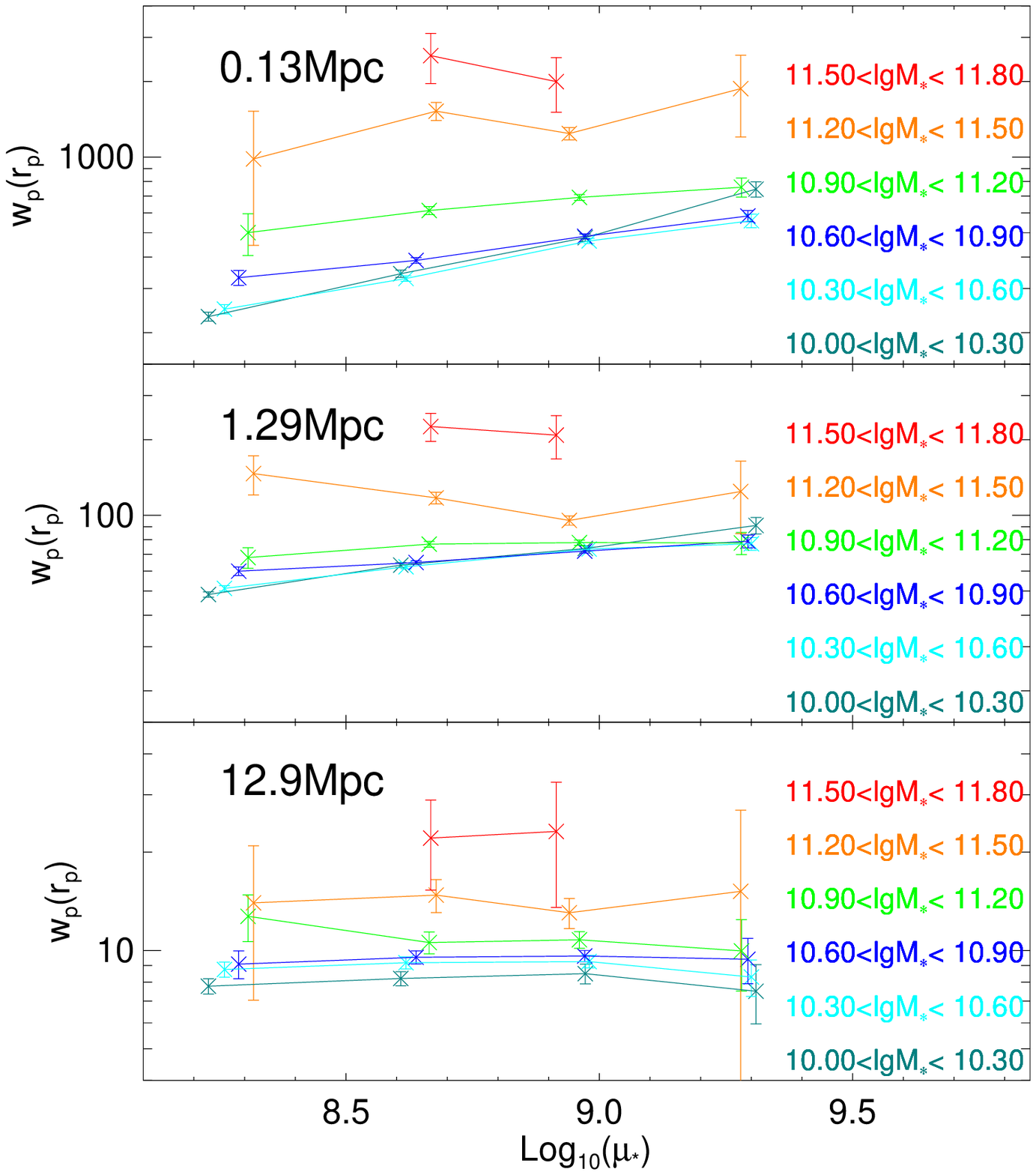,width=0.48\textwidth}
  \end{center}
    \caption{Projected cross-correlation function \wrp\ measured at
      different scales as indicated, as a function of \mstar\ for
      subsamples of different \mustar\ (left panels), and as a
      function of \mustar\ for subsamples  of different \mstar\ (right
      panels).}
  \label{fig:all_bias_mstar_mustar}
\end{figure*}

We further perform the same analysis to subsamples selected on the
plane of \mstar\ versus the surface stellar mass density (\mustar).
Results are displayed in Figure~\ref{fig:all_bias_mstar_mustar}. The
behaviors of \wrp\ on these figures are different from those on the
previous figures. In most cases the clustering amplitude shows
dependence only on stellar mass.  There is no  clear dependence on
\mustar\ in all cases except at small scales with $r_p$ below a few
Mpc and at low stellar masses with \lgmstar$\la 11$ (see the top two
panels on the left-hand),  where the clustering amplitude at fixed
scale and mass increases with increasing \mustar. 

%%%%%%%%%%%%%%%%%%%%%%%%%%%%%%%%%%%%%%%%%%%%%%%%%%

We have analyzed the clustering properties for both the central
galaxies in our sample and all the galaxies as a whole. For
completeness we have also measured \wrp\ for the satellite galaxies, by
excluding the central galaxies from our sample and selecting
subsamples of  satellites on the same two-dimensional planes as
above.  We find that, overall, the \wrp\ measured for the satellite galaxies
depends on the galaxy properties in a way that is qualitatively quite
similar to the \wrp\ measured above for the full galaxy population,
although for satellite galaxies we have to be limited to a smaller
number of subsamples with more noisy measurements in some cases, due
to the smaller number of satellite galaxies available in our sample.

The results in this section combine to reveal that, for galaxies as a
whole, the clustering amplitude depends on both stellar mass and other
properties considered, but in different ways for different types of
galaxies.  For massive galaxies with large \mstar\  (\mstar$\ga
10^{11}$M$_\odot$), which also have large \vdisp, \gr\ and \mustar,
the clustering is determined mainly on stellar mass,  
with little apparent dependence
on other properties  when \mstar\ is fixed.  At lower \mstar\, the
clustering depends on \vdisp\ and \gr\ at all scales in similar ways,
and on  \mustar\ only at scales smaller than a few Mpc.
In any case, the clustering amplitude shows little dependence on
\mstar\ at lower masses, when one of the other properties is fixed.
These results hold true when only considering satellite galaxies.
This implies that the different clustering properties at lower masses,
as seen in the central galaxy sample and the full sample, are mainly
caused by the satellite galaxies which dominate the clustering at the
low masses.

%%%%%%%%%%%%%%%%%%%%%%%%%%%%%%%%%%%%%%%%%%%%%%%%%%%%%%%%%%%%%%%%%%%%%
\section{Summary and Discussion}
\label{sec:summary}

In this paper we have studied the clustering of both central and
satellite galaxies in a large sample of $\sim$16,000  galaxy groups
identified by \citet{Yang-07} from the SDSS/DR7,  as well as the
clustering of the full population including both centrals and
satellites. In order to understand the co-dependence of clustering on
stellar mass (\mstar) and other properties, we have divided our
galaxies into subsamples on two-dimensional planes each formed by
\mstar\ and one of the other properties including stellar  velocity
dispersion (\vdisp), optical color (\gr) and surface stellar mass
density (\mustar). For each subsample we then measure the projected
cross-correlation functions \wrp\ with respect to a reference sample
consisting of about half a million galaxies, also selected from the
SDSS/DR7.  For each subsample of central galaxies, we have further
estimated an average dark matter halo mass (\mhalo), by modelling the
redshift distortion in the redshift-space cross-correlation function
between the centrals and the reference galaxies. We have compared
these measurements for the different populations, and for subsamples
selected by different properties. 

Our main conclusions can be summarized as follows.
\begin{enumerate}
\item For central galaxies, we  find that both the clustering
  amplitude on scales larger than a few Mpc and the dark matter halo
  mass show strongest dependence on \mstar, and there is no clear
  dependence on other properties when \mstar\ is fixed.  This result
  holds at all the stellar masses considered (\mstar$>2\times
  10^{10}$M$_\odot$). Our finding provides strong support to the
  commonly-adopted assumption that, for central galaxies, stellar mass
  is the galaxy property that is best indicative of the host dark halo
  mass.
\item The full galaxy population and the subset of satellites show
  similar clustering properties in all cases, which are similar to
  those of the central galaxies only at high-mass (\mstar$\ga
  10^{11}$M$_\odot$).
\item At low-mass (\mstar$\la 10^{11}$M$_\odot$), the clustering
  properties of the full population and the satellite galaxies differ
  from those of the central galaxies: the clustering amplitude
  increases with both \vdisp\ and \gr\ when \mstar\ is fixed, and
  depends very weakly on \mstar\ when \vdisp\ or \gr\ is fixed. 
\item At fixed \mstar, the surface mass density \mustar\ shows weak
  correlations with clustering amplitude (and halo mass in case of
  central galaxies). This is true for both centrals  and satellites,
  and holds at all masses  and on all scales, except that at $r_p$ below
  a few Mpc and \mstar$\la 10^{11}$M$_\odot$, the galaxies with higher
  surface density are more clustered.
\end{enumerate}

The strong mass dependence of galaxy clustering at the high mass end
has been well established in previous studies through measuring the
auto-correlation function of galaxies in SDSS and other surveys
\citep[e.g.][]{Tegmark-04, Norberg-01, Zehavi-04, Li-06b}. The same
effect is revealed in this work by the cross-correlation functions
which depend strongly on stellar mass for galaxies more massive than
$\sim 10^{11}$M$_\odot$. It is interesting that the effect holds even
when the galaxies are limited to narrow ranges in other properties.
More interestingly, we find that the same effect at high-mass is seen
when we divide the galaxies into centrals and satellites. It is  clear
that, for high-mass galaxies, stellar mass is the galaxy property that
is most strongly correlated with clustering.  This result was missing
in the previous study by \citet{Wake-Franx-vanDokkum-12} who measured
the \wrp\  for the full galaxy population as a function of both
\mstar\ and \vdisp,  using the same statistic and similar data as in
the current work. In order to understand the reason for the different
results in Wake et al. and here, we have performed an additional
analysis and present the results in the appendix
(\S\ref{sec:comp_wake}).  It is shown that the lack of mass dependence
at fixed \vdisp\ in Wake  et al. can be attributed to the too broad
mass  binning adopted,  which smoothed out the mass dependence at the
massive end.

At masses below $\sim 10^{11}$M$_\odot$, the clustering shows weak or
no dependence on stellar mass, also in agreement with previous studies
of the auto-correlation function of galaxies, and there is clear
dependence on both \vdisp\ and \gr\ on all scales probed even when
\mstar\ is fixed. The dependence on \vdisp\ at fixed mass found
here is apparently consistent with \citet{Wake-Franx-vanDokkum-12}. The similarity in the 
results for \vdisp\ and \gr\ may be reflecting the fact revealed by recent studies that
\vdisp\ is the best indicator of galaxy color when compared to other
properties such as stellar mass, surface mass density and morphology
\citep[e.g.][]{Wake-vanDokkum-Franx-12}. The result for \mustar\ is
somewhat different from that for \vdisp\ and \gr, in the sense that
the clustering amplitude shows dependence on \mustar\ only on  scales
smaller than a few Mpc, and this is true only for low-mass  galaxies
with \mstar$\la 10^{11}$M$_\odot$. The lack of dependence on
\mustar\  on large scales is again consistent with previous studies of
auto-correlation functions \citep{Li-06b}, which are found to be
independent of \mustar\ on scales above a few Mpc. Our result
apparently agrees with early studies that examined the correlation of
local environment with galaxy properties, finding color to be the
galaxy property most predictive of the local environment and
structural parameters to be almost independent of  local density at
fixed stellar mass \citep[e.g.][]{Blanton-05c, Kauffmann-04}. 

In L13 we concluded that, when compared to stellar velocity
dispersion, the stellar mass is better for indicating the host dark
halo mass of central galaxies. In this work we have extended this work
by including \gr\ and \mustar, which are the galaxy parameters
sensitive to recent star formation history and internal structure.  We
find that stellar mass is still the galaxy property that is most
tightly correlated with dark halo mass (and clustering). Therefore, it
is clear that the stellar mass  is better than other properties such
as stellar velocity dispersion,  optical color and surface mass
density, in terms of linking central galaxies  with their host
halo. One may question the reliability of taking the most massive
galaxy as the central  galaxy of groups \citep[e.g.][]{Skibba-11} . However, a recent study of
X-ray  observations of nearby clusters of galaxies
\citep{vonderLinden-14} has nicely shown that the typical offset is
only 20 kpc between the brightest cluster galaxy (BCG)  and X-ray
centroid, and that measurements of weak lensing mass centered  at BCGs
agree well with those centered at X-ray centroids. This suggested that
the BCGs (and similarly the most massive galaxy) are indeed very close
to the potential center of clusters. 

The full sample and the satellite population show similar clustering
properties, which are similar to those of central galaxies only at
high masses  (\mstar$\ga 10^{11}$M$_\odot$). This confirms our
previous conjecture that it is necessary to consider central and
satellite  galaxies separately when studying the link between galaxy
and dark  matter halos. The current work shows that the clustering of
the two types of galaxies depends on the galaxy properties in
different ways. Therefore, the behavior of clustering for a sample
including both populations must strongly depends on the
central/satellite fraction of the sample. It is thus not surprising
that the stellar mass is always most related with clustering and halo
mass at the massive end, where the sample is dominated by central
galaxies. At lower masses, in contrast, the satellite fraction is
larger and the result of the full sample is mainly reflecting the
behavior of satellite galaxies. 

Finally, we would like to point out that our results may be helpful
for understanding the role of dark matter halos in quenching the star
formation in central galaxies, a current topic that has been studied
in depth by many authors, both observationally and theoretically.  On
one hand, theoretically the so-called ``halo quenching'' process is
expected to be at work for central galaxies  of massive halos reaching
a critical halo mass of $\sim10^{12}$M$_\odot$, where the infalling
gas is shock-heated to the virial temperature and is no longer able to
cool efficiently \citep[e.g.][]{Silk-77, Rees-Ostriker-77}. On the
other hand, recent studies of the correlations between galaxy color
and structural parameters based on SDSS and high-z surveys have
reached a well-established conclusion that the presence of a prominent
central object such as a bulge is a necessary, but not sufficient
condition for quenching the star formation in central galaxies
\citep[e.g.][]{Kauffmann-06, Bell-08, Franx-08, Bell-12, Cheung-12,
  Fang-13}. This result supports the ``morphology quenching'' process
proposed by \citet{Martig-09}, in which the buildup of a central
massive spheroid can effectively stabilize the gas disk from
gravitationally collapsing and  forming stars. Recently
\citet{Fang-13} suggested a ``two-step'' picture in which  both the
halo quenching and morphology quenching processes play important
roles.  However, the two processes could be essentially the same
process if \vdisp\ turns out to be most correlated with dark matter
halo mass. In that case, halo mass would be the driving parameter for both
the growth of the bulge within the galaxy (as quantified by \vdisp)
and the gas heating in the host halo. This possibility can be ruled out
according to our analysis which  clearly show that the halo mass is
more tightly correlated with \mstar\ than with \vdisp.  Our result
suggests that halo quenching and morphology quenching  are two
distinct processes,  driven by different physical processes and
parameters.

%%%%%%%%%%%%%%%%%%%%%%%%%%%%%%%%%%%%%%%%%%%
\acknowledgments

We're grateful to the referee for the helpful comments.
This work is supported by National Key Basic Research Program of China
(No. 2015CB857004), NSFC (Grant No. 11173045, 11233005, 11325314,
11320101002) and the Strategic Priority Research Program ``The
Emergence of Cosmological Structures'' of CAS (Grant No. XDB09000000).

Funding for  the SDSS and SDSS-II  has been provided by  the Alfred P.
Sloan Foundation, the Participating Institutions, the National Science
Foundation, the  U.S.  Department of Energy,  the National Aeronautics
and Space Administration, the  Japanese Monbukagakusho, the Max Planck
Society,  and the Higher  Education Funding  Council for  England. The
SDSS Web  Site is  http://www.sdss.org/.  The SDSS  is managed  by the
Astrophysical    Research    Consortium    for    the    Participating
Institutions. The  Participating Institutions are  the American Museum
of  Natural History,  Astrophysical Institute  Potsdam,  University of
Basel,  University  of  Cambridge,  Case Western  Reserve  University,
University of Chicago, Drexel  University, Fermilab, the Institute for
Advanced   Study,  the  Japan   Participation  Group,   Johns  Hopkins
University, the  Joint Institute  for Nuclear Astrophysics,  the Kavli
Institute  for   Particle  Astrophysics  and   Cosmology,  the  Korean
Scientist Group, the Chinese  Academy of Sciences (LAMOST), Los Alamos
National  Laboratory, the  Max-Planck-Institute for  Astronomy (MPIA),
the  Max-Planck-Institute  for Astrophysics  (MPA),  New Mexico  State
University,   Ohio  State   University,   University  of   Pittsburgh,
University  of  Portsmouth, Princeton  University,  the United  States
Naval Observatory, and the University of Washington.

%%%%%%%%%%%%%%%%%%%%%%%%%%%%%%%%%%%%%
\appendix

\begin{figure*}
  \begin{center}
    \epsfig{figure=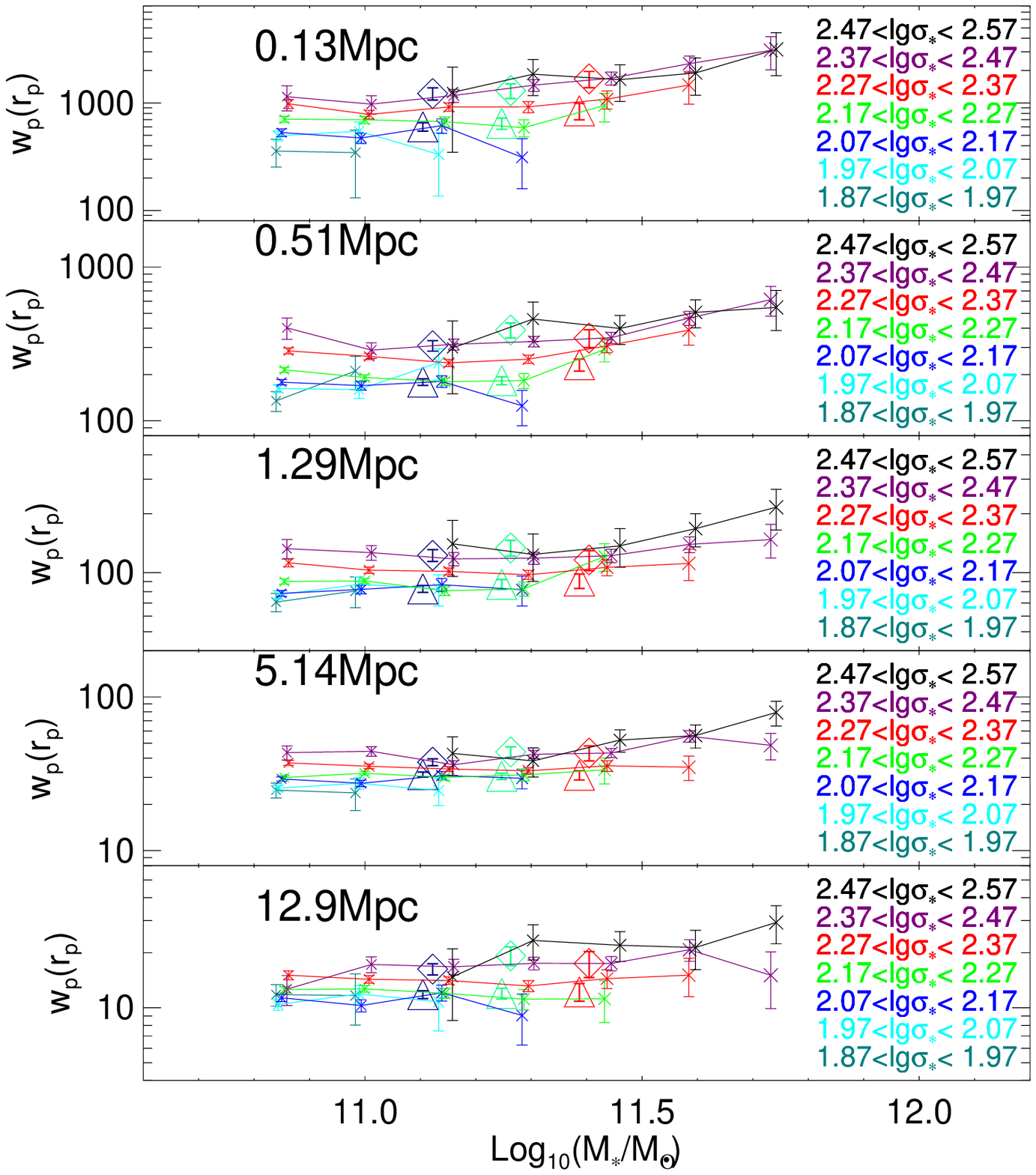,width=0.45\textwidth}
    \epsfig{figure=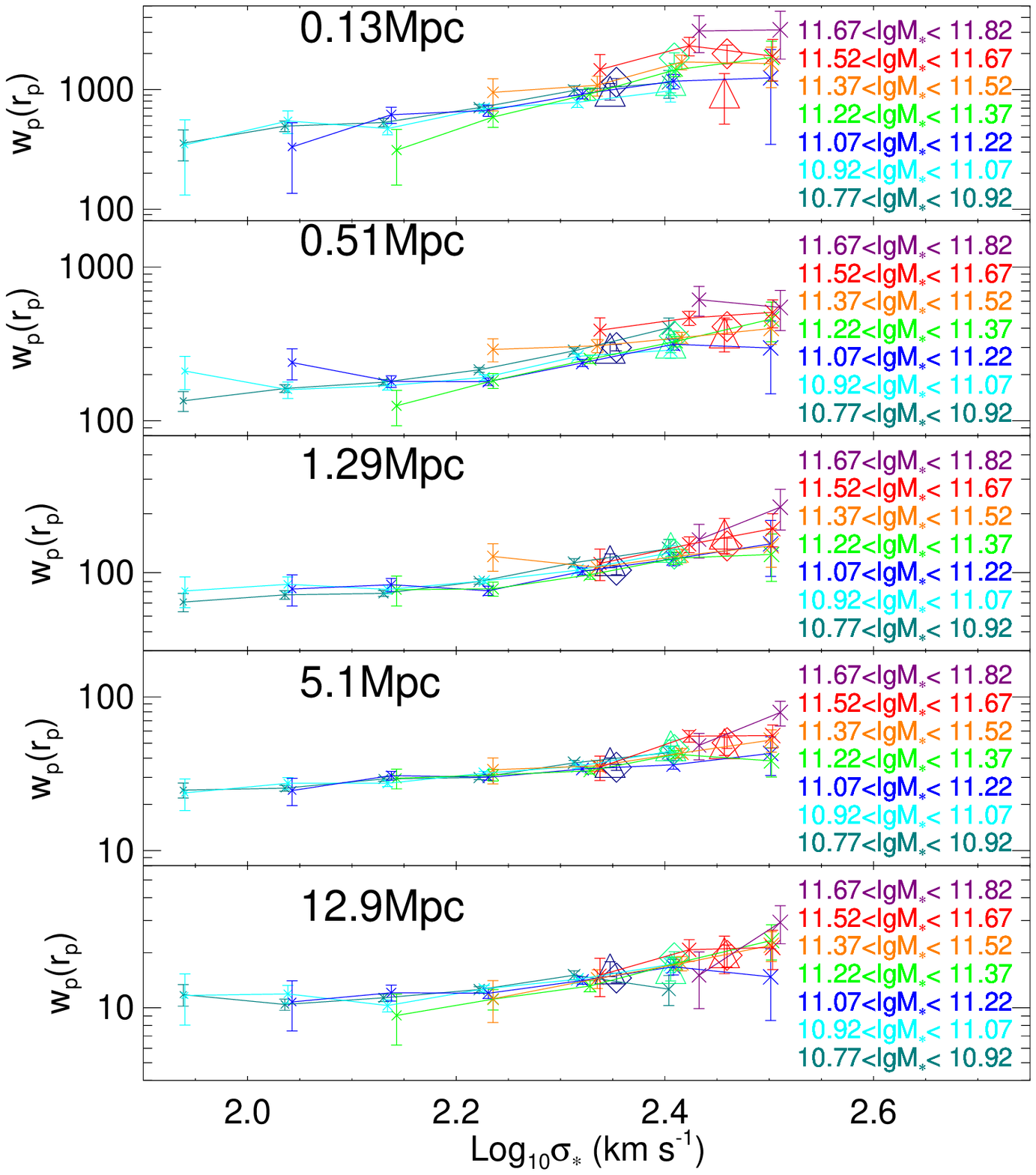,width=0.45\textwidth}
  \end{center}
  \caption{Same as the previous figure, except that we use stellar
    mass estimates and stellar velocity dispersion estimates from the
    MPA/JHU database following \citet{Wake-Franx-vanDokkum-12}, when
    dividing our galaxies into subsamples of \mstar\ and \vdisp.  Large
    triangles and diamonds present the results for subsamples selected with wider
    binning, adopting the same \mstar\ and \vdisp\ ranges as in
    \citet{Wake-Franx-vanDokkum-12}.}
  \label{fig:all_bias_wake}
\end{figure*}

\section{Examining the effect of different data sets and sample selections}
\label{sec:comp_wake}

Here we present an additional analysis which attempts to examine the
robustness of our results to the different definitions/estimations of
stellar mass and stellar velocity dispersion, as well as to understand
why the previous work by \citet{Wake-Franx-vanDokkum-12} led to
different conclusions. For the first purpose, we use \mstar\ and
\vdisp\ estimates from the MPA/JHU database,  the same data as used
by \citet{Wake-Franx-vanDokkum-12}, and we repeat the analysis as done
in the main text, selecting subsamples of the full galaxy population
on the two-dimensional plane of \mstar\ and \vdisp\ with exactly the
same binning scheme. The \wrp\ measurements are presented in
Figure~\ref{fig:all_bias_wake}.  Comparing this figure and
Figure~\ref{fig:all_bias_mstar_vdisp},  our conclusion remains
unchanged regarding the co-dependence of clustering on \mstar\ and
\vdisp,  although there are subtle differences in some cases.  Again,
we find the \wrp\ to depend mainly on \mstar\ at high masses and  on
\vdisp\ at low masses. 

Using the same data and the same statistic,
\citet{Wake-Franx-vanDokkum-12} reached a conclusion that is 
different from ours: there was no dependence of \wrp\ on stellar mass at fixed
\vdisp, but dependence on \vdisp\ was clearly seen at fixed \mstar.
However, the results from the two studies may not be comparable, as
Wake et al. adopted much wider binning when selecting their
subsamples.  In order to examine the effect of the different
selections, we follow  Wake et al. to select 12 subsamples on the
\mstar\ vs. \vdisp\ plane  with the same binning scheme.  The
\wrp\ measured at different scales as a function of the median
\mstar\ (left panels) or the median \vdisp\ (right panels) of these
subsamples are plotted in Figure~\ref{fig:all_bias_wake} as large
triangles and diamonds. The results are in good agreement with Wake et al.: the
clustering amplitude at given $r_p$ shows strong dependence on
\vdisp\ at fixed \mstar\ (see the comparison between triangles and diamonds 
in the left panels), and does not depend on \mstar\ at fixed \vdisp\ for scales beyond
$\sim0.5$ Mpc (see the comparison between triangles and diamonds in the bottom three panels at the
right hand). Comparing these subsamples with the subsamples selected
with finner binning, we find that the 6 subsamples in the left
panels are dominated by galaxies of relatively low masses
(\lgmstar$\la 11.5$), a regime where the  mass dependence of galaxy
clustering is weak as discussed above.  The mass dependence can only
be seen at the highest masses, and  can be easily hidden if the sample
being considered covers a wide  range of stellar mass.  Regarding the
6 subsamples in the right panels, we find that these subsamples are
biased to relatively large \vdisp, but none of them are representative
of  the galaxies with the highest \vdisp, only at which  the
clustering amplitude shows dependence on \mstar\ at fixed \vdisp.
Therefore, our analysis here explains why Wake et al. failed to detect
a mass dependence at fixed stellar velocity dispersion, which can be
simply attributed to the too broad binning in both \mstar\ and
\vdisp\ adopted in their sample selection.

%%%%%%%%%%%%%%%%%%%%%%%%%%%%%%%%%%%%%

%%%%%%%%%%%%%%%%%%%%%%%%%%%%%%%%%%%%%%%%%%%%%%%%%%%%%

%%%%%%%%%%%%The End%%%%%%%%%%%%%%%%%%%%%%%%%%%%%%%%%%%%%%%%%
\label{lastpage}
\end{document}